\documentclass[fleqn,usenatbib]{mnras}

\usepackage{newtxtext,newtxmath}

\usepackage[T1]{fontenc}

\DeclareRobustCommand{\VAN}[3]{#2}
\let\VANthebibliography\thebibliography
\def\thebibliography{\DeclareRobustCommand{\VAN}[3]{##3}\VANthebibliography}

\usepackage{graphicx}	
\usepackage{amsmath}	

\title[Formation of compact relics.]{The formation of compact massive relic galaxies in MOND}

\author[Eappen et al.]{ Robin Eappen$^{1}$\thanks{E-mail: eappenr@sirrah.troja.mff.cuni.cz} and Pavel Kroupa$^{1,2}$\thanks{E-mail: pkroupa@uni-bonn.de} 
\\
$^{1}$Charles University in Prague, Faculty of Mathematics and Physics, Astronomical Institute, V Holešovickách 2, CZ-180 00 Praha 8, Czech Republic\\
$^{2}$Helmholtz-Institut für Strahlen- und Kernphysik (HISKP), Universität Bonn, Nussallee 14–16, 53115 Bonn, Germany}

\date{Accepted 2024 January 23. Received 2024 January 13; in original form 2022 August 12}

\pubyear{2024}

\begin{document}
\label{firstpage}
\pagerange{\pageref{firstpage}--\pageref{lastpage}}
\maketitle

\begin{abstract}
Compact massive relic galaxies are a class of galaxies that exhibit characteristics suggesting they have remained largely unchanged since their initial formation, making them "relics" of the early universe. These galaxies represent a distinct class characterized by strongly peaked high-velocity dispersion profiles with large rotational velocities. This study investigates the formation of such galaxies within the framework of Milgromian Dynamics (MOND), offering a unique perspective on their origin without invoking the presence of cold or warm dark matter. Our analysis focuses on the collapse dynamics of isolated non-rotating post-Big-Bang gas clouds, revealing kinematic and density profiles comparable to observed compact massive relic galaxies like NGC 1277, Mrk 1216, and PGC 032873. The findings underscore the natural emergence of compact massive relic galaxies within a MOND-based Universe, providing valuable insights into the interplay between gravitational dynamics and galaxy formation.
\end{abstract}

\begin{keywords}
galaxies: evolution -- galaxies: formation -- galaxies: kinematics and dynamics -- software: simulations.
\end{keywords}



\section{Introduction}
\label{sec:Introduction}

The most accepted form of cosmological model, known as the dark-energy cold-dark-matter ($\Lambda$CDM) model, posits that galaxies form within the cosmic web, influenced by gravitational forces and quantum fluctuations during the early universe (\citealt{2020A&A...641A...6P,2005pfc..book.....M}). This model is understood to explain observed features of the universe's large-scale structure. The $\Lambda$CDM model suggests that early-type galaxies (ETGs) observed in the present day formed through the size growth of the initially compact quiescent galaxies found at higher redshifts (\citealt{2005ApJ...626..680D, 2007ApJ...665..265F, 2009ApJ...697.1290B, 2010ApJ...709.1018V, 2014ApJ...788...28V}).

The two-phase galaxy formation paradigm further details this evolution. In the "cosmic dawn," gravitational instabilities lead to the emergence of protogalactic systems (\citealt{1980lssu.book.....P,2010gfe..book.....M}). Computational simulations, like those by \citet{2002Sci...295...93A}, support the role of these instabilities. The subsequent "cosmic noon" witnesses heightened star formation and galaxy maturation through mergers and accretion (\citealt{2015ARA&A..53...51S, 2006ApJS..163....1H}; but see \citealt{2023MNRAS.524.3252H}). Observational studies, including those from the Hubble Space Telescope and surveys like the Sloan Digital Sky Survey, contribute essential insights, allowing a nuanced understanding of galaxy evolution across cosmic epochs (\citealt{2016ApJ...830...83C,2014MNRAS.441..599B}). This two-phase model integrates observations and simulations to illuminate the complex interplay between cosmic dynamics and astrophysical mechanisms, offering a comprehensive view of galaxies' diverse properties from their formation to the present universe.

\citet{2009ApJ...699L.178N}, using a high-resolution $\Lambda$CDM hydrodynamical cosmological simulation, illustrates the potential existence of compact and massive elliptical galaxies at high redshifts. Ongoing investigations, including recent observations from the James-Web-Space-Telescope (JWST), have contributed to the exploration of this phenomenon \citep{2023Natur.619..716C}. The simulation by \citet{2009ApJ...699L.178N} reveals that the accretion of stripped stellar material and the dynamical friction of surviving stellar cores contribute to the increase in galaxy size and the reduction of central concentration over time. The overall study suggests that minor mergers, involving the accretion of weakly bound material, play a crucial role in the late-stage evolution of early-type galaxies, allowing compact high-redshift spheroids to transform into present-day systems resembling observed local ellipticals. \citet{2010ApJ...725.2312O}, in their study of ETGs, show that the early phase (redshift z $>$ 2) of the two-phase process involves in-situ star formation within the galaxy from infalling cold gas, followed by an extended phase (z $<$ 2) dominated by the accretion of ex-situ stars using $\Lambda$CDM cosmological simulations. Their study concludes that the majority of in-situ stars are formed at z > 2 from cold gas flows, and the observational result of "archaeological downsizing" is recovered, where the oldest stars are found in the centers of most massive galaxies.

In $\Lambda$CDM, the progenitors of ETGs undergo wet mergers accompanied by intense star formation, resulting in the rapid growth of stellar mass while maintaining compact sizes, yielding the observed compact massive galaxies. These resulting galaxies undergo dry mergers, explaining the increase in stellar size and other structural and chemical evolution aspects of massive ETGs (\citealt{2005ApJ...626..680D}). Galaxies resembling high-redshift compact quiescent galaxies, when discovered in the nearby Universe, are referred to as relic galaxies \citep{2009ApJ...692L.118T,2009ApJ...695..101D}. These compact massive relic galaxies are objects that formed in the early phase of the Universe and are thought to be unaltered since their formation (\citealt{2017MNRAS.467.1929F, 2024MNRAS.527.8793S}). \citet{2014ApJ...780L..20T} suggest that these compact massive relic galaxies formed their stars rapidly indicating they have short star-formation timescales (SFTs) like early-type galaxies (ETGs, \citealt{2005ApJ...621..673T,2005ApJ...632..137N,2015MNRAS.448.3484M,2021A&A...655A..19Y}). Their star formation history (SFH) would peak at the beginning of the Universe and would not have any significant star formation later on.

Discussing these compact relics has not been easy as originally there were no well-defined criteria to identify them (i.e concerning the limits of compactness and masses). A fully detailed study of the structural and stellar population properties for a sample of such galaxies by \citet{2009ApJ...692L.118T} rejected all candidates after a detailed analysis of their morphologies, structural profiles, kinematics and stellar populations. Examining the kinematics, morphology, and stellar properties of nearby compact massive galaxies revealed that these nearby objects exhibit elongated morphologies, fast rotation, young luminosity-weighted ages, and recent significant star formation bursts, with some being compatible with having very old stellar populations, distinguishing them as unique entities that deviate from the typical patterns observed in present-day massive ellipticals, spirals, or dwarfs \citep{2012MNRAS.423..632F, 2012ApJ...751...45T}. NGC 1277 was claimed to be the first fully confirmed massive relic \citep{2014ApJ...780L..20T}. Subsequently, a dozen new massive relic candidates with similar properties to NGC 1277 and hosting extreme super-massive black holes were reported \citep{2012Natur.491..729V}. Comapct relic galaxies are a peculiar type of object when compared to their massive elliptical galaxy counter parts as the former are observed to have large velocity dispersions and large rotational velocities whereas elliptical galaxies are observed to be more 'pressure-supported' systems (i.e. with lower rotational velocities, \citealt{2014ApJ...780L..20T,2017MNRAS.467.1929F}).

Over the last decade, a cluster of putative relic galaxies has been identified in close proximity (z $\approx$ 0.05) and subsequently validated for their relic classification through meticulous examinations of diverse attributes, including galaxy-wide initial mass functions (IMF), globular cluster populations, and dynamical properties (\citealt{2014ApJ...780L..20T, 2015ApJ...808...79F,2017MNRAS.467.1929F,2015MNRAS.452.1792Y,2017MNRAS.468.4216Y,2021A&A...654A.136S,2019MNRAS.487.4939M,2021ApJ...914...20K}). The ongoing INSPIRE survey endeavors to corroborate additional relic candidates at redshifts extending up to z = 0.5 \citep{2021A&A...654A.136S}. The criteria delineating relic galaxy candidates mirror attributes reminiscent of high-redshift red nuggets, necessitating substantial mass ($\approx$ $10^{12} M_{\odot}$), compactness ($r_{\rm eff}$ $\leq$ 1.5 kpc), quiescence, and notable age (age > 10 Gyr). While observational investigations into relic galaxies in the nearby Universe furnish valuable insights into broader processes of galaxy formation and evolution, the direct observation of their individual formative processes remains an elusive pursuit.

Recent $\Lambda$CDM cosmological simulations explore the evolution of individual compact galaxies over time, without the constraints and biases inherent in real observations (\citealt{2015MNRAS.449..361W,2015MNRAS.449.2396S,2016MNRAS.458.2371R,2017MNRAS.465..722F}). However, these simulations also possess limitations, emphasizing the importance of analog searches within simulations as a means to assess and refine numerical models (\citealt{2015MNRAS.446..521S,2016MNRAS.463.3948D,2018MNRAS.474.3976G,2018MNRAS.477.1206N,Pillepich_2018,2019MNRAS.486.2827D,2020NatRP...2...42V}).

Although the $\Lambda$CDM cosmological simulations are able to reproduce the compact massive relic galaxies and achieve "archaeological downsizing" (but have not yet been able to address how and why massive galaxies form quicker in the early Universe compared to the less massive galaxies) it is in tension with the observed fraction of elliptical galaxies in the Universe. \citet{2010A&A...509A..78D} finds that there has not been a significant change in the fraction of elliptical galaxies compared to what it was 6 Gyr ago (see also \citealt{2014A&A...570A.102T}). \citet{2018MNRAS.475.3700M} find that most stars in elliptical galaxies formed over 10 Gyr ago. \citet{2020NatAs...4..252S} analyse nearly 29000 galaxies from the Baryon Oscillation Spectroscopic Survey finding that the residual star formation contributed less than one percent mass fraction  in massive ETGs over the past 2Gyr, with a decreasing trend in young star fraction with increasing galaxy mass. \citet{2022MNRAS.515.4514S} in their study suggest an intrinsic, in situ process triggering star formation at later epochs in NGC 1277, emphasizing a general constraint on the presence of young stars in the cores of massive early-type galaxies, with potential contributions from evolved stars. \citet{2016MNRAS.463.3409V} show that spheroids predominantly consist of old stars, and \citet{2011MNRAS.418L..74D} and \citet{2015MNRAS.448.3484M} demonstrate faster and earlier star formation in massive ETGs, resulting in compact, metal-rich, and $\alpha$-enhanced systems.
\citet{2022ApJ...925..183H} highlights the tension between the observed proportion of galaxy morphological types and the predictions of the $\Lambda$CDM framework, indicating the need for further investigation and potential refinements to the current understanding of galaxy formation and dynamics. 

Here, we report the formation of compact massive relic galaxies to be naturally arising in a monolithic post Big-Bang non-rotating cloud collapse in a Milgromian Dynamics (MOND) simulation. We caution that we are not performing any direct comparison with compact relic galaxies formed in $\Lambda$CDM as our work is solely based on galaxies formed in the MOND framework. Also, a direct comparison to results from $\Lambda$CDM simulations may be considered of limited interpretative value as our work is not based on cosmological simulations. Interested readers can see 
\citet{2015MNRAS.449..361W,2016MNRAS.456.1030W}, \citet{2016MNRAS.460.1147B}, \citet{2019MNRAS.485..396V} for more details on the formation of compact galaxies in $\Lambda$CDM. In previous work, we have shown (\citealt{2022MNRAS.516.1081E}) that the SFTs observed for ETGs are a natural occurrence in MOND, but found some discrepancies in structure and morphology, possibly related to the neglect of early mergers. It, therefore, appeared that our simulated galaxies should rather resemble compact massive relic galaxies, which we demonstrate hereafter.

Regular ETGs, in particular ellipticals, could be the product of a monolithic collapse and the subsequent merger of similar-mass galaxies. That is, a proto-massive group or cluster forms from a collapsing massive post-Big-Bang gas cloud with the inner cMpc (co-moving Mpc) region forming two or more early-type galaxies (each forming as in \citealt{2022MNRAS.516.1081E}) that then merge to become the present-day ETGs. Here, we revisit one non-merger model from the suit of simulations of \citet{2022MNRAS.516.1081E} which is in agreement with the downsizing and also aligns with recent findings from JWST observations on rapid early galaxy formation (\citealt{2023Natur.619..716C, 2022arXiv220712446L}) to show in more depth that such models agree extremely well with the observed relic galaxies which are assumed to have remained largely unevolved since their formation (\citealt{2009ApJ...692L.118T,2014ApJ...780L..20T, 2017MNRAS.467.1929F, 2024MNRAS.527.8793S}

Section \ref{sec:two} discusses the kinematical and density properties of the observed and simulated compact massive relic galaxies and Section \ref{sec:three} contains the concluding remarks. 

\section{Are compact relics Naturally arising in MOND?}
\label{sec:two}

A correction to the Newtonian theory of gravity, MOND, was proposed by \citet{1983ApJ...270..365M} to account for the ubiquitous mass discrepancies in the Universe, without invoking the exotic dark matter that is required if one adheres to standard dynamics. MOND introduces a new constant, $a_{\rm 0}$, into physics which has the dimensions of acceleration and the principle of MOND is that the standard form of dynamics shifts to space-time scale invariance of the equation of motion under the transformation $(t,r) \xrightarrow{} (\lambda t, \lambda r)$, where $\lambda$ is an arbitrary number, when the acceleration is much smaller than $a_{\rm 0}$ \citep{2009ApJ...698.1630M}. This new dynamics can be formulated as a non-relativistic gravitational theory by generalising the classical action for gravity: the extremum of the action then yields a generalised Poisson equation (see \citealt{1984ApJ...286....7B,2008arXiv0801.3133M,2010MNRAS.403..886M,2014SchpJ...931410M,2012LRR....15...10F,2022Symm...14.1331B} for detailed reviews on the theory). This theory is found to accommodate the observed laws of galactic dynamics such as asymptotically flat rotation curves and also predicted galaxy scaling relations obeyed by galaxies such as the Baryonic Tully Fisher Relation \citep{2000ApJ...533L..99M,2005ApJ...632..859M,2012AJ....143...40M} and the Radial Acceleration Relation \citep{1990A&ARv...2....1S,2004ApJ...609..652M,2017ApJ...836..152L}.

Structure formation in a MOND cosmology was pioneered by \citet{1998MNRAS.296.1009S}. He proposed that the thermal and dynamical history of the early MOND Universe is exactly that of the standard Big-Bang model and all predictions relevant to the nucleosynthesis of the light elements carry over to MOND cosmology and after non-relativistic matter dominates the mass density of the Universe. Due the the non-linearity of gravitation, structure formation in MOND cosmology diverges from that of standard cosmology. This is also implied by \citet{2002MNRAS.331..909N} who notes the rapid growth of small fluctuations with MOND. The \citet{2009ApJ...694.1220M} study provides insights into the dynamics of structure formation and evolution within a MOND framework, highlighting the accelerated formation of structures without the need for dark matter. This picture from the spherical collapse model in MOND provides a bottom-top scenario for the structure formation in which the smaller structures formed before the larger ones. The cosmological structure formation simulations in the $\nu$HDM MOND model \citep{2023MNRAS.523..453W} simulations demonstrate that the formation of a cosmic web of filaments and voids is not unique to standard Einstein/Newton-based cosmology, indicating that MOND-based cosmological frameworks can also reproduce this characteristic feature of the large-scale structure in the universe. This however does not mean that a MOND cosmological model automatically solves all problems, as for example, the $\nu$HDM model appears to form galaxies too late and also too massive structures overall \citep{2023arXiv230911552K}.

The actual formation of galaxies with simulations of dissipationless collapse in MOND was investigated for the first time by \citet{2007ApJ...660..256N} and \citet{2008MNRAS.386.1588S}. This work shows less-massive galaxies would collapse before the more massive ones (fig. 3 in \citealt{1998MNRAS.296.1009S}, fig. 4 in \citealt{2008MNRAS.386.1588S}) suggesting inconsistency with the observed downsizing. The work presented in \citet{2022MNRAS.516.1081E} re-addresses the formation of elliptical galaxies using hydro-dynamical MOND simulations with star formation. Given that the background MOND cosmological model is not yet clearly defined (e.g. \citealt{2013ApJ...772...10K} and \citealt{2021PhRvL.127p1302S}), the work presented here follows the collapse of post-Big-Bang gas clouds with no rotation. \citet{2022MNRAS.516.1081E} extract the required conditions for the formed ETGs to confer with downsizing, finding that the gas clouds need to start with a radius near 500 kpc independently of mass.

A numerical simulation technique is needed to solve the equations of motion of particles and the hydro-dynamical equations with star-formation and baryonic feedback algorithms that describe the mathematical MOND models to allow to study galaxy formation. The Phantom of Ramses code developed in Bonn (POR, \citealt{2015CaJPh..93..232L,2021arXiv210111011N} a customized version of RAMSES, \citealt{2002A&A...385..337T}) uses a multi-grid code to solve the generalised MOND Poisson equation\footnote{\citet{2015MNRAS.446.1060C} developed a similar code independently called RAyMOND.}. \citet{2022MNRAS.516.1081E} use POR to constrain the initial conditions of the post-Big-Bang gas clouds required to form ETGs and we use one of the model galaxies described in that work for the study here. We note that the properties of observed disc galaxies arise from the collapse of rotating gas clouds \citep{2020ApJ...890..173W}. The employed star-formation and feedback sub-grid physics adequately reproduce observed disc galaxies \citep{2022MNRAS.tmp.3410N}.

\begin{figure}
	\includegraphics[width=\columnwidth]{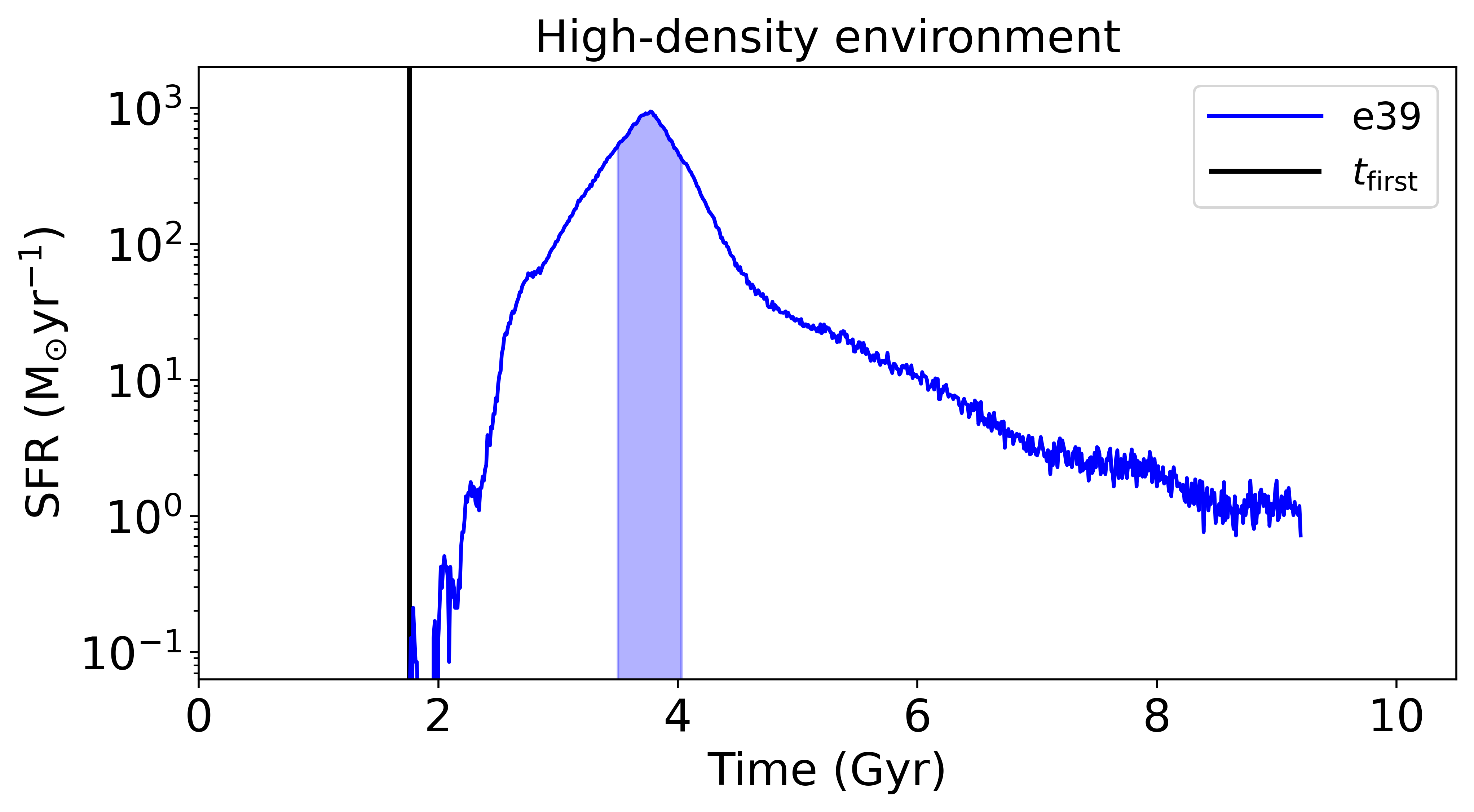}
    \caption{SFH of the model galaxy formed in MOND (Table \ref{tab:one}). The solid black line is the time when the first stellar particle is formed and the shaded blue region is the star-forming timescale (SFT), $\Delta\tau_{\rm m}$. $t$ = 0 corresponds to the Big-Bang.}
    \label{fig:SFHEmond}
\end{figure}

\begin{table}
	\caption{Initial conditions and properties of the model galaxy from \citet{2022MNRAS.516.1081E}.}
	\label{tab:one}.
\begin{tabular}{c c c c c c}
   
		\hline
		Model & $M_{\rm initial}$ & $R_{\rm initial}$ & $\Delta\tau_{\rm m}$ & $r_{\rm eff}$ & $\it{SFR}_{\rm peak}$\\
		Name & (10$^{11}$ $M_\odot$) & (kpc) & (Gyr) & (kpc) & ($M_{\odot}\rm yr^{-1}$)\\
        \hline
e39 & 0.7 & 500 & 0.54 & 1.04 & 9.39E+02 \\
        \hline
    \end{tabular}
\\Note: Column 1: Model name, Column 2 ($M_{\rm initial}$): initial mass of the post-Big-Bang gas cloud, Column 3 ($R_{\rm initial}$): initial radius of the post-Big-Bang gas cloud. Column 4 ($\Delta\tau_{\rm m}$): SFT. Column 5 ($r_{\rm eff}$): projected effective-radius and Column 6 ($\it{SFR}_{\rm peak}$): SFR at the peak of the SFH. SFT is defined as the full - width at half maximum of the distribution of the ages of stellar particles (see Fig. \ref{fig:SFHEmond}).
\end{table}

That  ETGs may have acquired significant amount of mass through later mergers of star-forming galaxies (as expected in $\Lambda$CDM theory) is disfavoured by the NUV and optical survey of \citet{2021MNRAS.500.3368S} and by the comparison of the CO absorption features in NGC 1277 with a large representative sample of ETGs (\citealt{2022MNRAS.515L..56E}). The observed ETGs thus appear to have formed the vast majority of their stellar mass rapidly on a time scale comparable to the downsizing time-scale of \citet{2005ApJ...621..673T}, \citet{2015MNRAS.448.3484M} (see also \citealt{2021A&A...655A..19Y}). \citet{2022MNRAS.516.1081E} show that the SFTs observed (as deduced by \citealt{2005ApJ...621..673T}) for ETGs are a natural occurrence in the MOND paradigm if ETGs form from monolithically collapsing non-rotating post-Big-Bang gas clouds. This quick and early mass assembly is very similar to the compact massive relics as described in \citet{2014ApJ...780L..20T}, \citet{2017MNRAS.467.1929F} and \citet{2024MNRAS.527.8793S}. 

Here we use the model galaxy 'e39' from \citet{2022MNRAS.516.1081E} (Table \ref{tab:one}) to compare its kinematics and density profile with real compact massive relics in the Universe (Table \ref{tab:values}). The model galaxy is formed through a monolithic collapse of a non-rotating gas cloud of initial mass 0.7 $\times$ 10$^{11} M_{\odot}$ and initial radius of 500 kpc. The simulations effectively assume the canonical stellar IMF \citep{2001MNRAS.322..231K} for the formation of stellar particles. Only the simple cooling/heating feedback algorithm is used as it is entirely adequate (\citealt{2022MNRAS.tmp.3410N}) and the size of the simulation box is 1000 kpc which achieves a maximum of 0.24 kpc and minimum of 7.81 kpc in spatial resolution (see \citealt{2022MNRAS.516.1081E} for more details on the initial conditions of the gas cloud and simulation parameters). We use the data published in \citet{2014ApJ...780L..20T} and \citet{2017MNRAS.467.1929F} for the galaxies NGC 1277, PGC 032873 and Mrk 1216. \citet{2022MNRAS.516.1081E} apply the average stellar age-mass relation for ETGs deduced by \citet{2005ApJ...621..673T} to place their model galaxies in a time frame relative to the Big-Bang. This is done by shifting the peak of the SFH of a model galaxy to the average age deduced by \citet{2005ApJ...621..673T} for a galaxy with the same mass. This allows \citet{2022MNRAS.516.1081E} to calculate the time when the first stellar particles would have formed in the model galaxies relative to the Big-Bang. The SFH of the model galaxy listed in Table \ref{tab:one} is plotted in Fig. \ref{fig:SFHEmond} which gives us the time when the first stellar particle formed in the model galaxy, $t_{\rm first}$ = 1.75 Gyr after Big-Bang in a high-density environment and $t_{\rm first}$ = 3.95 Gyr after the Big-Bang in a low-density environment (see \citealt{2022MNRAS.516.1081E} for more details). This is similar to compact massive relics (\citealt{2014ApJ...780L..20T, 2017MNRAS.467.1929F, 2024MNRAS.527.8793S}) where NGC 1277, PGC 032873 and Mrk 1216 are found to have formed their stars around 10 Gyr ago. Detailed chemical enrichment histories of such collapsing post-Big-Bang gas clouds by applying a self-consistently evolving galaxy-wide IMF are available in \citet{2021A&A...655A..19Y}. Concerning the age-dating applied in Fig. \ref{fig:SFHEmond}: the peak of the SFH was gauged in \citet{2022MNRAS.516.1081E} using the age-dating by \citet{2005ApJ...621..673T} which has an uncertainty of about 1 Gyr. It is thus likely that some "e39"-type relics would have finished forming by 3 Gyr after the Big-Bang, as suggested by \citet{2024MNRAS.527.8793S}.

\begin{table}
	\caption{Structural properties of the compact relics.}
	\label{tab:values}
\begin{tabular}{c c c c c}
   
		\hline
		Galaxy & $r_{\rm eff}$ & $V_{\rm \sigma}$ & $V_{\rm rot}$ & Mass\\
		Name & (kpc) & (km/s) & (km/s) &(10$^{11}$ $M_\odot$)\\
        \hline
NGC 1277 & 1.2 $\pm$ 0.1 & 385 $\pm$ 6 & 292 & 1.3 $\pm$ 0.3 \\
PGC 032873 & 1.8 $\pm$ 0.2 & 358 $\pm$ 5 & 285 & 2.3 $\pm$ 0.9 \\
Mrk 1216 & 2.3 $\pm$ 0.1 & 368 $\pm$ 3 & 200 & 2.0 $\pm$ 0.8 \\
e39 & 1.04 & 369.5 & 203 & 0.7 \\

        \hline
    \end{tabular}
\\Note: Column 1: Name, Column 2: effective-radius, Column 3: central velocity dispersion, Column 4: rotational velocity at $r_{\mathrm{eff}}$ and Column 5: mass of the galaxy. The values are taken from \citet[see their Fig. 2]{2017MNRAS.467.1929F}
\end{table}

\subsection{Kinematical properties}

\begin{figure}
	\includegraphics[width=\columnwidth]{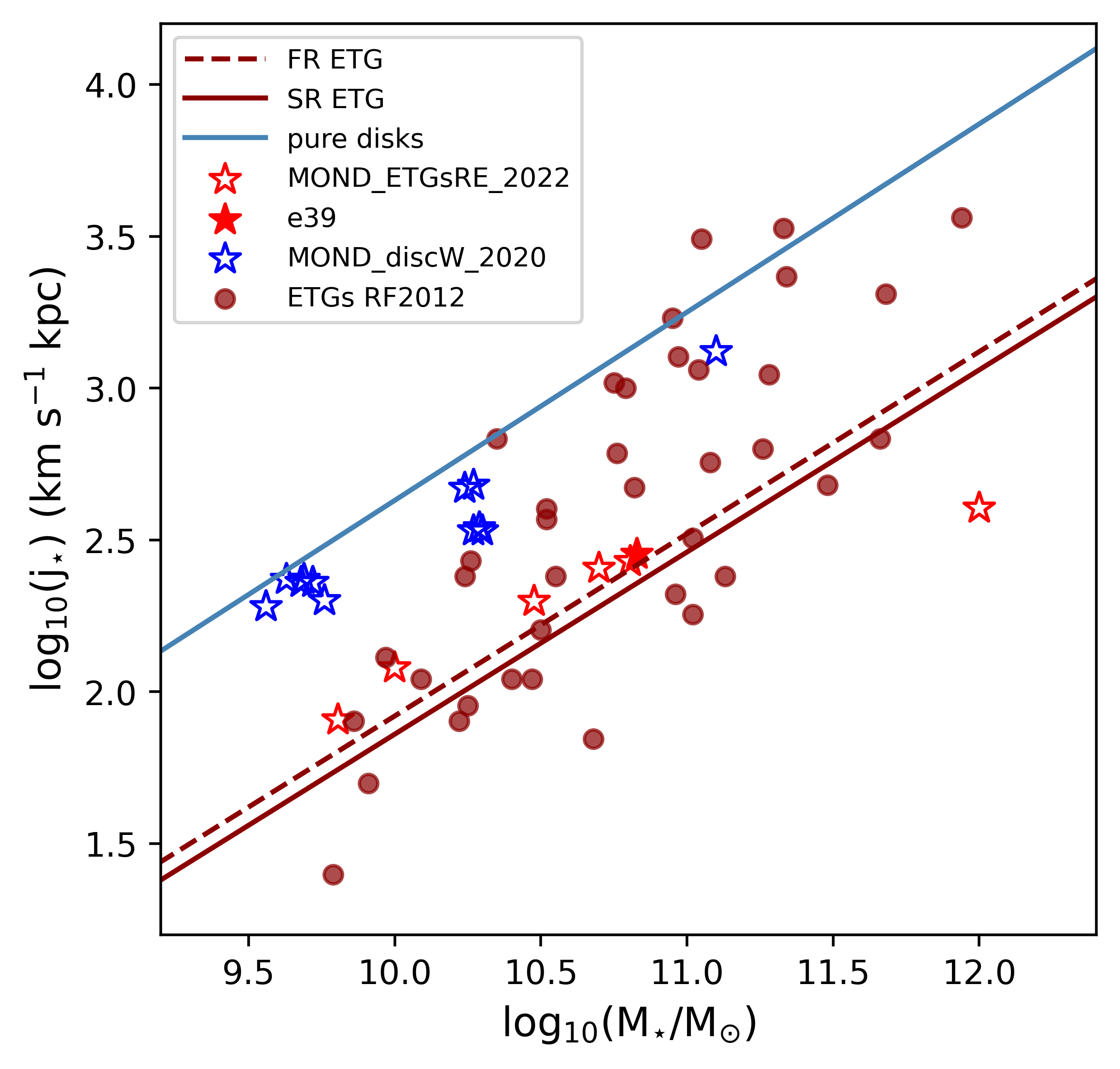}
    \caption{Classification diagram of galaxies, using the parameter space of stellar mass and specific angular momentum. The solid blue line shows the scaling relation for pure disc galaxies and the dashed red and solid red line shows, respectively, the scaling relation for fast- and slow-rotating elliptical galaxies (\citealt{2012ApJS..203...17R}). The red filled circles are the observed ETGs classified in terms of their morphology (ranging from S0 to E5, see table 5 of \citealt{2012ApJS..203...17R}). The blue open stars are the model disc galaxies from \citet{2020ApJ...890..173W} and the red open stars are the model ETGs from \citet{2022MNRAS.516.1081E}. The red filled star is the model galaxy e39.}
    \label{fig:angmom}
\end{figure}

\begin{figure}
	\includegraphics[width=\columnwidth]{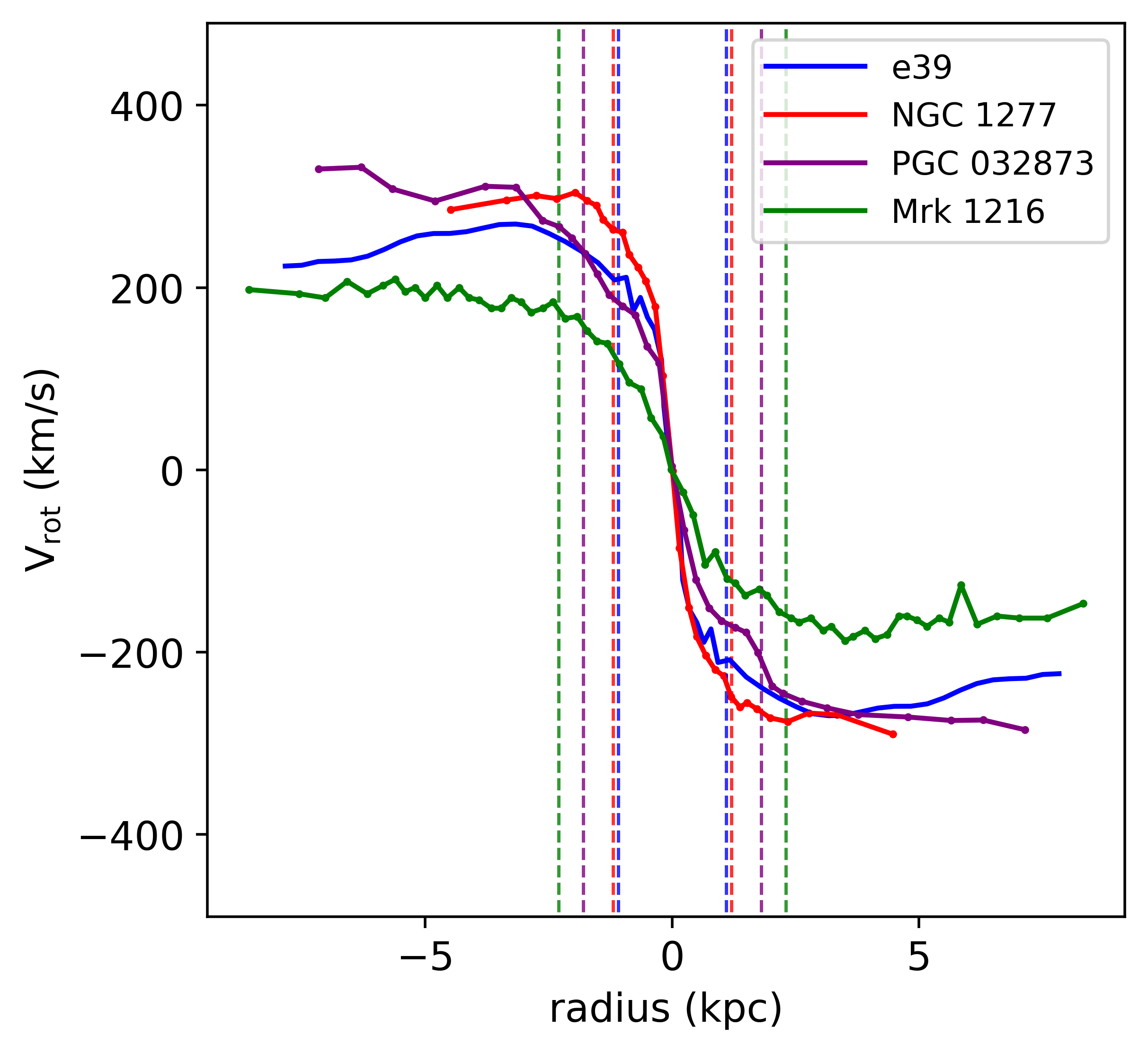}
    \caption{Rotational velocity profile of the ETG formed in MOND (e39) compared with observed compact massive relic galaxies \citep{2017MNRAS.467.1929F} listed in Table \ref{tab:values}. The dashed lines show the effective radii of the galaxies.}
    \label{fig:vcirc}
\end{figure}

\begin{figure}
	\includegraphics[width=\columnwidth]{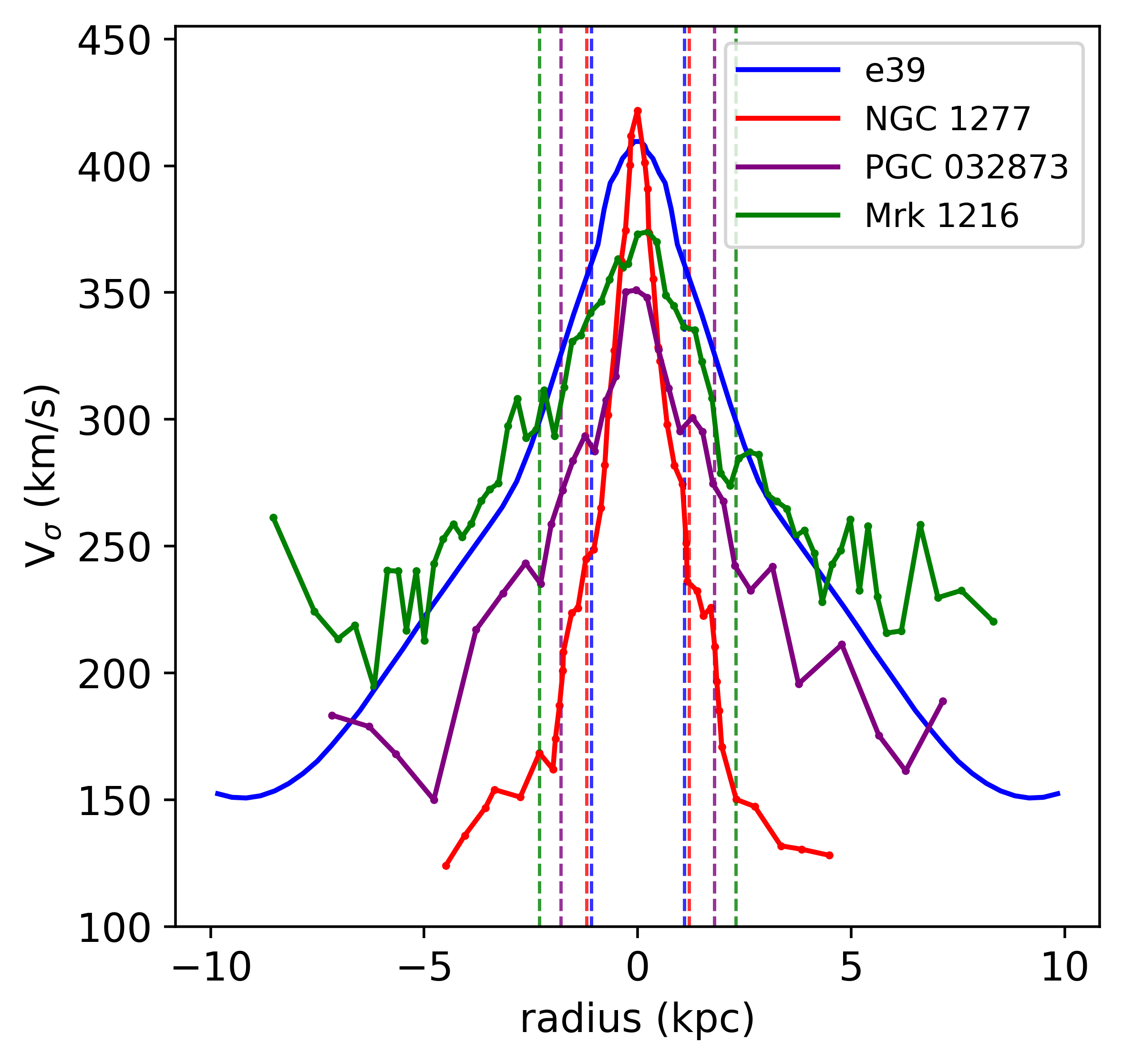}
    \caption{Velocity dispersion profile of the ETG formed in MOND (e39) compared with observed compact massive relic galaxies \citep{2017MNRAS.467.1929F} listed in Table \ref{tab:values}. The dashed lines show the effective radii of the galaxies.}
    \label{fig:veldisp}
\end{figure}

\begin{table}
	\caption{The specific angular momentum properties of the model galaxy.}
	\label{tab:angmomprop}
\begin{tabular}{c c c c c}
   
		\hline
		Model & $j_{\rm x}$ & $j_{\rm y}$ & $j_{\rm z}$ & $j_{\rm \star}$\\
		Name & ($\rm{km \ s^{-1} \ kpc}$) & ($\rm{km \ s^{-1} \ kpc}$) & ($\rm{km \ s^{-1} \ kpc}$) & ($\rm{km \ s^{-1} \ kpc}$)\\
        \hline
        e39 & 1.65 & -5.39 & -284.01 & 284.06 \\
        \hline
    \end{tabular}
\\Note: $j_{\rm{x}}$, $j_{\rm{y}}$ and $j_{\rm{z}}$ are the components of the specific angular momentum vector of the model galaxy along the x, y and the z directions respectively and $j_{\rm \star}$ is the absolute value of the specific angular momentum.
\end{table}

The specific angular momentum 
is computed (Appendix \ref{sec:appendix2}) and Table \ref{tab:angmomprop} lists the angular momentum properties of the model galaxy. The direction of the rotation for this model is along the z-axis. The specific angular momentum of the model galaxy is consistent with the specific angular momentum -- mass scaling relation deduced by \citet{2012ApJS..203...17R} for early-type galaxies which puts the model into the category of fast-rotating ETG (Fig. \ref{fig:angmom}). The structural decomposition of fast-rotating ETGs uncovers that there is a hidden disc component that provides the additional angular momentum (\citealt{2012ApJS..203...17R}). Fig. \ref{fig:angmom} shows that the model galaxies formed through the monolithic collapse of rotating gas clouds in MOND fall on the observed angular momentum -- mass relation for late-type galaxies \citep{2020ApJ...890..173W}, while the model galaxies formed through monolithic collapse of non-rotating gas clouds in MOND \citep{2022MNRAS.516.1081E} fall on the observed angular momentum -- mass relation for ETGs. A detailed study of the morphology and other structural properties of these model galaxies will be discussed in a follow-up paper (Eappen et al. in preparation).

The rotational velocity along the z-direction (Fig. \ref{fig:vcirc}), $v_{\mathrm{rot}}$, is calculated as a function of radius, $r$ of the model galaxy. The velocity dispersion of the model galaxy is calculated by averaging the face-on projection of line-of-sight velocity dispersion of all the stellar particles of the model galaxy (similar to \citealt{2017MNRAS.467.1929F} in their observational study, see Appendix \ref{sec:appendix1}).

The compact massive relic galaxies are observed to have large rotational velocities and a high central velocity dispersion. They are very fast rotators compared to elliptical galaxies with comparable mass. The model galaxy e39 from \citet{2022MNRAS.516.1081E} is comparable to all the three compact massive relics in terms of the effective radius ($r_{\rm eff}$) and mass.

The velocity dispersion profiles of these galaxies are shown in Fig. \ref{fig:veldisp}. It can be seen that the model galaxy has a similar central peak for the velocity dispersion profile to that of the observed compact massive relics but here the decline is not as steep as for NGC 1277. This is likely due to the external field effect since NGC1277 is embedded in a galaxy cluster while model e39 is isolated (the extrernal field effect is a part of MOND that arises because of its non-linear nature and can be ignored for studies involving isolated models, see \citealt{1983ApJ...270..365M,2022Symm...14.1331B} for more details). The rotational velocity profile (Fig. \ref{fig:vcirc}) for the model galaxy also follows the same trend as the observed compact massive relics but has a lower velocity at $r_{\rm eff}$ than NGC 1277. The unique peaked kinematics exhibited by these compact massive relic galaxies may stem from the existence of super-massive black holes, potentially initiating and evolving independently or at varying rates relative to the evolutionary pathways of their host galaxies during earlier cosmic epochs (\citealt{2012Natur.491..729V,2015ApJ...808...79F, 2016ApJ...817....2W, 2020MNRAS.498.5652K, 2023ApJ...958..186C}).

Although the initial cloud has no angular momentum, the forming galaxy spins up. This is a consequence of the collapse of the cloud not proceeding exactly spherically symmetrically due to not perfect initial spherical symmetry in the numerical model which amplifies due to the enhanced gravitation in MOND and angular momentum conservation. This leads to off-center formation of stellar particles which leads to spherically non-symmetrical gas flows, such that the final compact galaxy always rotates. In the real Universe, gas clouds will never be without angular momentum because of tidal torquing and turbulence, and the models here study the rare collapse of post-Big-Bang clouds that have a negligible specific angular momentum. The more typical outcome of the collapse of rotating clouds leads to disc galaxies with exponential surface density profiles \citep{2020ApJ...890..173W}.

\subsection{Stellar mass density profile}

\begin{figure}
	\includegraphics[width=\columnwidth]{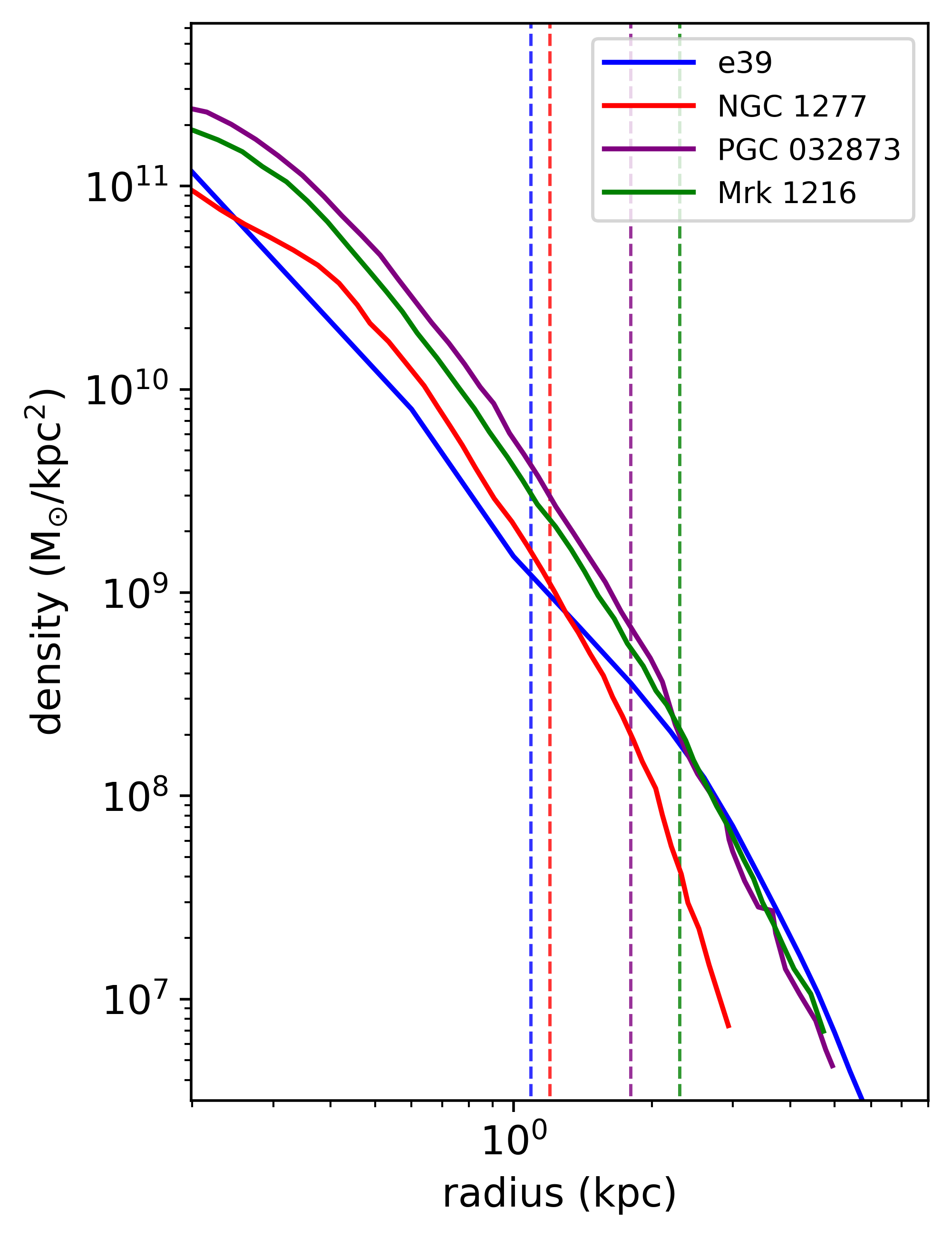}
    \caption{Surface density profile of the ETG formed in MOND (e39) compared with observed compact massive relic galaxies \citep{2017MNRAS.467.1929F}. The dashed lines show the effective radii of the galaxies.}
    \label{fig:denprof}
\end{figure}

\citet{2014ApJ...780L..20T} and \citet{2017MNRAS.467.1929F} show that the compact massive relics have different stellar mass density profiles compared to present-day disc or elliptical galaxies. When compared to the densest elliptical galaxies, the density profile of NGC 1277 is found to have a denser profile within $r_{\rm eff}$ being under-dense beyond $r_{\rm eff}$. Compact massive relic galaxies are found to be equivalent to massive galaxies with a similar stellar mass found in the early Universe \citep{2014ApJ...780L..20T}. The stellar mass density profile of the model galaxy e39 is similar to the stellar mass density profile of the compact massive relic galaxies as shown in Fig. \ref{fig:denprof}.

\section{Conclusions}
\label{sec:three}

If the most massive galaxies are formed from one or two major mergers of compact galaxies of comparable masses then this would mean that the intrinsic properties, such as the SFT of a galaxy, would not change significantly over time through mergers and it is the structural properties like $r_{\rm eff}$ and morphology that would passively evolve over time. The model galaxies from \citet{2022MNRAS.516.1081E} were found to follow the downsizing behaviour as noted in \citet{2005ApJ...621..673T,2010MNRAS.404.1775T, 2021A&A...655A..19Y} and formed under monolithic collapse of non-rotating post-Big-Bang gas clouds. Here we study the kinematical properties of one of the models from \citet{2022MNRAS.516.1081E} of comparable mass with the observed compact massive relic galaxies NGC 1277, PGC 032873 and Mrk 1216 \citep{2014ApJ...780L..20T,2017MNRAS.467.1929F} listed in Table \ref{tab:values} and find the model to be very similar to these observed compact massive relic galaxies. These compact massive relics are found to have their average stellar ages similar to ETGs of comparable mass in a high-density environment as found by \citet{2005ApJ...621..673T}, where a galaxy with $M_{*} = 10^{11} M_{\odot}$ would have formed 10 Gyr ago. We find that,

\begin{itemize}
    \item The model galaxy has a short SFT (i.e. it forms early and rapidly, \citealt{2022MNRAS.516.1081E}) similar to compact massive relic galaxies. The model galaxy forms most of its stellar particles around 4 Gyr after the Big-Bang (assuming \citealt{2005ApJ...621..673T} age-dating, see Fig. \ref{fig:SFHEmond}) but falls short to the observed compact massive relics (which forms most of its stellar content within 2 Gyr after the Big-Bang, see \citealt{2017MNRAS.467.1929F}). Allowing for the observed dispersion of the ages of ETGs by \citet{2005ApJ...621..673T}, however, avoids this problem.
    \item The effective-radius ($r_{\rm eff}$ $\approx 1$ kpc) of the model galaxy is comparable to NGC 1277 which has a similar mass ($M_{*} \approx$ 10$^{11} M_{\odot}$).
    \item The model galaxy has a high central velocity dispersion similar to the compact massive relics in \citet{2017MNRAS.467.1929F} (Fig. \ref{fig:veldisp}).
    \item The model galaxy and other ETGs in MOND (\citealt{2022MNRAS.516.1081E}) are found to fall onto the observed angular momentum -- mass relation for ETGs which classes them as fast rotators (Fig. \ref{fig:angmom}) with a rotational velocity profile that is comparable to the observed compact massive relics (Fig. \ref{fig:vcirc}).
    \item The stellar mass density profile of the model galaxy is steep, similarly to the compact massive relics (Fig. \ref{fig:denprof}).
\end{itemize}


\citet{2023A&A...675A.143C} argue that the relic NGC 1277 cannot be understood within the standard dark-matter based cosmological structure formation models due to its apparent lack of dark matter, while the regular massive ETG NGC 1278 can. For a model of galaxy formation to be valid, it needs to account for all galaxies equally well. While in MOND galaxies "lacking dark matter" can be obtained if the galaxy is immersed in an external field that effectively lowers its phantom dark matter halo, self-consistent cosmological structure formation simulation are needed to directly address the likelihood of occurrence of such a case.

It is quite remarkable that a post-Big-Bang non-rotating gas cloud collapse in MOND yields a galaxy that is a fast rotator with very high central velocity dispersion similar to the properties of the observed compact massive relic galaxies. Recent work by \citet{2024MNRAS.527.8793S} shows that most of the stellar content in the compact relic galaxies would have formed before 3 Gyr since the Big-Bang. A direct comparison between the ages of the observed galaxies with our models in this work is not directly possible, as our approach does not rest on cosmological simulations. Instead, we rely on ages from \citet{2005ApJ...621..673T} to identify an approximate timeframe for the formation of our models in the real Universe. Quasar activity only 0.7 Gyr after the Big Bang as well as the recent discovery of very high redshift galaxies (reaching almost $\mathrm{10^{10}}$ $M_{\rm \odot}$ only about 0.5 Gyr after the Big Bang, see \citealt{2023Natur.619..716C,2022arXiv220712446L, 2022ApJ...939L..31H}) indicate that galaxies could have formed even faster in the real Universe, which could perhaps be obtained if some post-Big-Bang gas clouds are denser (compared to the model galaxy e39) with a smaller initial radius, $R_{\rm initial}$. Our result shows that the Milgromian law of gravitation naturally yields objects resembling real ones through the monolithic collapse of gas clouds, which did not have to be the case.

While some studies show recent starbursts in compact massive galaxies, these are nevertheless dominated by very old stars \citep{2012MNRAS.423..632F}. In fact, \citet{2022MNRAS.515.4514S} find a 0.8 per cent mass fraction of young stars in the centre of NGC 1277, similar to that found in massive ETGs. This suggests that the low levels of young stellar populations in the cores of massive ETGs could be from self-regulated gas returned to the interstellar medium from stellar evolution \citep{2022MNRAS.515.4514S}. The collapse models studied here are thus consistent with the major property of relics in them being compact, fast rotating with high velocity dispersion and of very old age.

The monolithic collapse, incidentally, also naturally explains the rapid emergence of quasars and super-massive black holes and their correlation with host-galaxy-properties \citep{2020MNRAS.498.5652K}. According to these authors, the monolithic collapse first forms an extremely massive star-burst cluster at the centre. This star burst cluster has a top-heavy IMF due to the low metallicity and high density and may appear like a quasar \citep{2017A&A...608A..53J}. The cluster shields itself \citep{2019MNRAS.484.4379P} from further in-falling gas from the forming galaxy through the quasar-like radiation field generated by the ionising stars. After about 50 Myr these die leaving a very large population of stellar-mass black holes. The now in-falling gas from the forming galaxy causes the black hole cluster to shrink to a relativistic condition at which point it implodes catastrophically to a super-massive black hole of near final mass through the radiation of gravitational waves. According to the conservative calculations by \citet{2020MNRAS.498.5652K}, this whole process takes less time than the time-scale for the formation of the ETG. The correlation of the so formed super-massive black hole mass with the ETG mass comes naturally through the integrated galaxy-wide IMF (IGIMF, \citealt{2003ApJ...598.1076K,2006MNRAS.365.1333W,2018A&A...620A..39J,2021A&A...655A..19Y}) theory that essentially accounts for the mass budget in stars and star clusters in terms of the star-formation rate of the galaxy.

\section*{Acknowledgements}
The numerical simulations were performed using the Tiger--Cluster of the Astronomical Institute of Charles University, Prague. R.E. is supported by the Grant Agency of Charles University under grant No. 234122. We thank the DAAD Eastern European grant at Bonn University for supporting the research visits. We would also like to thank the referee for their valuable input which helped in improving the manuscript.

\section*{Data Availability}
 
All data used here have been generated as described using the publicly available POR code and the cited literature.


\bibliographystyle{mnras}
\bibliography{example} 

\begin{thebibliography}{}
\makeatletter
\relax
\def\mn@urlcharsother{\let\do\@makeother \do\$\do\&\do\#\do\^\do\_\do\%\do\~}
\def\mn@doi{\begingroup\mn@urlcharsother \@ifnextchar [ {\mn@doi@} {\mn@doi@[]}}
\def\mn@doi@[#1]#2{\def\@tempa{#1}\ifx\@tempa\@empty \href {http://dx.doi.org/#2} {doi:#2}\else \href {http://dx.doi.org/#2} {#1}\fi \endgroup}
\def\mn@eprint#1#2{\mn@eprint@#1:#2::\@nil}
\def\mn@eprint@arXiv#1{\href {http://arxiv.org/abs/#1} {{\tt arXiv:#1}}}
\def\mn@eprint@dblp#1{\href {http://dblp.uni-trier.de/rec/bibtex/#1.xml} {dblp:#1}}
\def\mn@eprint@#1:#2:#3:#4\@nil{\def\@tempa {#1}\def\@tempb {#2}\def\@tempc {#3}\ifx \@tempc \@empty \let \@tempc \@tempb \let \@tempb \@tempa \fi \ifx \@tempb \@empty \def\@tempb {arXiv}\fi \@ifundefined {mn@eprint@\@tempb}{\@tempb:\@tempc}{\expandafter \expandafter \csname mn@eprint@\@tempb\endcsname \expandafter{\@tempc}}}

\bibitem[\protect\citeauthoryear{{Abel}, {Bryan}  \& {Norman}}{{Abel} et~al.}{2002}]{2002Sci...295...93A}
{Abel} T.,  {Bryan} G.~L.,   {Norman} M.~L.,  2002, \mn@doi [Science] {10.1126/science.295.5552.93}, \href {https://ui.adsabs.harvard.edu/abs/2002Sci...295...93A} {295, 93}

\bibitem[\protect\citeauthoryear{{Banik} \& {Zhao}}{{Banik} \& {Zhao}}{2022}]{2022Symm...14.1331B}
{Banik} I.,  {Zhao} H.,  2022, \mn@doi [Symmetry] {10.3390/sym14071331}, \href {https://ui.adsabs.harvard.edu/abs/2022Symm...14.1331B} {14, 1331}

\bibitem[\protect\citeauthoryear{{Barber}, {Schaye}, {Bower}, {Crain}, {Schaller}  \& {Theuns}}{{Barber} et~al.}{2016}]{2016MNRAS.460.1147B}
{Barber} C.,  {Schaye} J.,  {Bower} R.~G.,  {Crain} R.~A.,  {Schaller} M.,   {Theuns} T.,  2016, \mn@doi [\mnras] {10.1093/mnras/stw1018}, \href {https://ui.adsabs.harvard.edu/abs/2016MNRAS.460.1147B} {460, 1147}

\bibitem[\protect\citeauthoryear{{Bekenstein} \& {Milgrom}}{{Bekenstein} \& {Milgrom}}{1984}]{1984ApJ...286....7B}
{Bekenstein} J.,  {Milgrom} M.,  1984, \mn@doi [\apj] {10.1086/162570}, \href {https://ui.adsabs.harvard.edu/abs/1984ApJ...286....7B} {286, 7}

\bibitem[\protect\citeauthoryear{{Bezanson}, {van Dokkum}, {Tal}, {Marchesini}, {Kriek}, {Franx}  \& {Coppi}}{{Bezanson} et~al.}{2009}]{2009ApJ...697.1290B}
{Bezanson} R.,  {van Dokkum} P.~G.,  {Tal} T.,  {Marchesini} D.,  {Kriek} M.,  {Franx} M.,   {Coppi} P.,  2009, \mn@doi [\apj] {10.1088/0004-637X/697/2/1290}, \href {https://ui.adsabs.harvard.edu/abs/2009ApJ...697.1290B} {697, 1290}

\bibitem[\protect\citeauthoryear{{Bluck}, {Mendel}, {Ellison}, {Moreno}, {Simard}, {Patton}  \& {Starkenburg}}{{Bluck} et~al.}{2014}]{2014MNRAS.441..599B}
{Bluck} A. F.~L.,  {Mendel} J.~T.,  {Ellison} S.~L.,  {Moreno} J.,  {Simard} L.,  {Patton} D.~R.,   {Starkenburg} E.,  2014, \mn@doi [\mnras] {10.1093/mnras/stu594}, \href {https://ui.adsabs.harvard.edu/abs/2014MNRAS.441..599B} {441, 599}

\bibitem[\protect\citeauthoryear{{Candlish}, {Smith}  \& {Fellhauer}}{{Candlish} et~al.}{2015}]{2015MNRAS.446.1060C}
{Candlish} G.~N.,  {Smith} R.,   {Fellhauer} M.,  2015, \mn@doi [\mnras] {10.1093/mnras/stu2158}, \href {https://ui.adsabs.harvard.edu/abs/2015MNRAS.446.1060C} {446, 1060}

\bibitem[\protect\citeauthoryear{{Carnall} et~al.,}{{Carnall} et~al.}{2023}]{2023Natur.619..716C}
{Carnall} A.~C.,  et~al., 2023, \mn@doi [\nat] {10.1038/s41586-023-06158-6}, \href {https://ui.adsabs.harvard.edu/abs/2023Natur.619..716C} {619, 716}

\bibitem[\protect\citeauthoryear{{Cohn} et~al.,}{{Cohn} et~al.}{2023}]{2023ApJ...958..186C}
{Cohn} J.~H.,  et~al., 2023, \mn@doi [\apj] {10.3847/1538-4357/ad029d}, \href {https://ui.adsabs.harvard.edu/abs/2023ApJ...958..186C} {958, 186}

\bibitem[\protect\citeauthoryear{{Comer{\'o}n} et~al.,}{{Comer{\'o}n} et~al.}{2023}]{2023A&A...675A.143C}
{Comer{\'o}n} S.,  et~al., 2023, \mn@doi [\aap] {10.1051/0004-6361/202346291}, \href {https://ui.adsabs.harvard.edu/abs/2023A&A...675A.143C} {675, A143}

\bibitem[\protect\citeauthoryear{{Conselice}, {Wilkinson}, {Duncan}  \& {Mortlock}}{{Conselice} et~al.}{2016}]{2016ApJ...830...83C}
{Conselice} C.~J.,  {Wilkinson} A.,  {Duncan} K.,   {Mortlock} A.,  2016, \mn@doi [\apj] {10.3847/0004-637X/830/2/83}, \href {https://ui.adsabs.harvard.edu/abs/2016ApJ...830...83C} {830, 83}

\bibitem[\protect\citeauthoryear{{Daddi} et~al.,}{{Daddi} et~al.}{2005}]{2005ApJ...626..680D}
{Daddi} E.,  et~al., 2005, \mn@doi [\apj] {10.1086/430104}, \href {https://ui.adsabs.harvard.edu/abs/2005ApJ...626..680D} {626, 680}

\bibitem[\protect\citeauthoryear{{Damjanov} et~al.,}{{Damjanov} et~al.}{2009}]{2009ApJ...695..101D}
{Damjanov} I.,  et~al., 2009, \mn@doi [\apj] {10.1088/0004-637X/695/1/101}, \href {https://ui.adsabs.harvard.edu/abs/2009ApJ...695..101D} {695, 101}

\bibitem[\protect\citeauthoryear{{Dav{\'e}}, {Angl{\'e}s-Alc{\'a}zar}, {Narayanan}, {Li}, {Rafieferantsoa}  \& {Appleby}}{{Dav{\'e}} et~al.}{2019}]{2019MNRAS.486.2827D}
{Dav{\'e}} R.,  {Angl{\'e}s-Alc{\'a}zar} D.,  {Narayanan} D.,  {Li} Q.,  {Rafieferantsoa} M.~H.,   {Appleby} S.,  2019, \mn@doi [\mnras] {10.1093/mnras/stz937}, \href {https://ui.adsabs.harvard.edu/abs/2019MNRAS.486.2827D} {486, 2827}

\bibitem[\protect\citeauthoryear{{Delgado-Serrano}, {Hammer}, {Yang}, {Puech}, {Flores}  \& {Rodrigues}}{{Delgado-Serrano} et~al.}{2010}]{2010A&A...509A..78D}
{Delgado-Serrano} R.,  {Hammer} F.,  {Yang} Y.~B.,  {Puech} M.,  {Flores} H.,   {Rodrigues} M.,  2010, \mn@doi [\aap] {10.1051/0004-6361/200912704}, \href {https://ui.adsabs.harvard.edu/abs/2010A&A...509A..78D} {509, A78}

\bibitem[\protect\citeauthoryear{{Dubois}, {Peirani}, {Pichon}, {Devriendt}, {Gavazzi}, {Welker}  \& {Volonteri}}{{Dubois} et~al.}{2016}]{2016MNRAS.463.3948D}
{Dubois} Y.,  {Peirani} S.,  {Pichon} C.,  {Devriendt} J.,  {Gavazzi} R.,  {Welker} C.,   {Volonteri} M.,  2016, \mn@doi [\mnras] {10.1093/mnras/stw2265}, \href {https://ui.adsabs.harvard.edu/abs/2016MNRAS.463.3948D} {463, 3948}

\bibitem[\protect\citeauthoryear{{Eappen}, {Kroupa}, {Wittenburg}, {Haslbauer}  \& {Famaey}}{{Eappen} et~al.}{2022}]{2022MNRAS.516.1081E}
{Eappen} R.,  {Kroupa} P.,  {Wittenburg} N.,  {Haslbauer} M.,   {Famaey} B.,  2022, \mn@doi [\mnras] {10.1093/mnras/stac2229}, \href {https://ui.adsabs.harvard.edu/abs/2022MNRAS.516.1081E} {516, 1081}

\bibitem[\protect\citeauthoryear{{Eftekhari}, {La Barbera}, {Vazdekis}  \& {Beasley}}{{Eftekhari} et~al.}{2022}]{2022MNRAS.515L..56E}
{Eftekhari} E.,  {La Barbera} F.,  {Vazdekis} A.,   {Beasley} M.,  2022, \mn@doi [\mnras] {10.1093/mnrasl/slac068}, \href {https://ui.adsabs.harvard.edu/abs/2022MNRAS.515L..56E} {515, L56}

\bibitem[\protect\citeauthoryear{{Faber} et~al.,}{{Faber} et~al.}{2007}]{2007ApJ...665..265F}
{Faber} S.~M.,  et~al., 2007, \mn@doi [\apj] {10.1086/519294}, \href {https://ui.adsabs.harvard.edu/abs/2007ApJ...665..265F} {665, 265}

\bibitem[\protect\citeauthoryear{{Famaey} \& {McGaugh}}{{Famaey} \& {McGaugh}}{2012}]{2012LRR....15...10F}
{Famaey} B.,  {McGaugh} S.~S.,  2012, \mn@doi [Living Reviews in Relativity] {10.12942/lrr-2012-10}, \href {https://ui.adsabs.harvard.edu/abs/2012LRR....15...10F} {15, 10}

\bibitem[\protect\citeauthoryear{{Ferr{\'e}-Mateu}, {Vazdekis}, {Trujillo}, {S{\'a}nchez-Bl{\'a}zquez}, {Ricciardelli}  \& {de la Rosa}}{{Ferr{\'e}-Mateu} et~al.}{2012}]{2012MNRAS.423..632F}
{Ferr{\'e}-Mateu} A.,  {Vazdekis} A.,  {Trujillo} I.,  {S{\'a}nchez-Bl{\'a}zquez} P.,  {Ricciardelli} E.,   {de la Rosa} I.~G.,  2012, \mn@doi [\mnras] {10.1111/j.1365-2966.2012.20897.x}, \href {https://ui.adsabs.harvard.edu/abs/2012MNRAS.423..632F} {423, 632}

\bibitem[\protect\citeauthoryear{{Ferr{\'e}-Mateu}, {Mezcua}, {Trujillo}, {Balcells}  \& {van den Bosch}}{{Ferr{\'e}-Mateu} et~al.}{2015}]{2015ApJ...808...79F}
{Ferr{\'e}-Mateu} A.,  {Mezcua} M.,  {Trujillo} I.,  {Balcells} M.,   {van den Bosch} R. C.~E.,  2015, \mn@doi [\apj] {10.1088/0004-637X/808/1/79}, \href {https://ui.adsabs.harvard.edu/abs/2015ApJ...808...79F} {808, 79}

\bibitem[\protect\citeauthoryear{{Ferr{\'e}-Mateu}, {Trujillo}, {Mart{\'\i}n-Navarro}, {Vazdekis}, {Mezcua}, {Balcells}  \& {Dom{\'\i}nguez}}{{Ferr{\'e}-Mateu} et~al.}{2017}]{2017MNRAS.467.1929F}
{Ferr{\'e}-Mateu} A.,  {Trujillo} I.,  {Mart{\'\i}n-Navarro} I.,  {Vazdekis} A.,  {Mezcua} M.,  {Balcells} M.,   {Dom{\'\i}nguez} L.,  2017, \mn@doi [\mnras] {10.1093/mnras/stx171}, \href {https://ui.adsabs.harvard.edu/abs/2017MNRAS.467.1929F} {467, 1929}

\bibitem[\protect\citeauthoryear{{Furlong} et~al.,}{{Furlong} et~al.}{2017}]{2017MNRAS.465..722F}
{Furlong} M.,  et~al., 2017, \mn@doi [\mnras] {10.1093/mnras/stw2740}, \href {https://ui.adsabs.harvard.edu/abs/2017MNRAS.465..722F} {465, 722}

\bibitem[\protect\citeauthoryear{{Genel} et~al.,}{{Genel} et~al.}{2018}]{2018MNRAS.474.3976G}
{Genel} S.,  et~al., 2018, \mn@doi [\mnras] {10.1093/mnras/stx3078}, \href {https://ui.adsabs.harvard.edu/abs/2018MNRAS.474.3976G} {474, 3976}

\bibitem[\protect\citeauthoryear{{Haslbauer}, {Banik}, {Kroupa}, {Wittenburg}  \& {Javanmardi}}{{Haslbauer} et~al.}{2022a}]{2022ApJ...925..183H}
{Haslbauer} M.,  {Banik} I.,  {Kroupa} P.,  {Wittenburg} N.,   {Javanmardi} B.,  2022a, \mn@doi [\apj] {10.3847/1538-4357/ac46ac}, \href {https://ui.adsabs.harvard.edu/abs/2022ApJ...925..183H} {925, 183}

\bibitem[\protect\citeauthoryear{{Haslbauer}, {Kroupa}, {Zonoozi}  \& {Haghi}}{{Haslbauer} et~al.}{2022b}]{2022ApJ...939L..31H}
{Haslbauer} M.,  {Kroupa} P.,  {Zonoozi} A.~H.,   {Haghi} H.,  2022b, \mn@doi [\apjl] {10.3847/2041-8213/ac9a50}, \href {https://ui.adsabs.harvard.edu/abs/2022ApJ...939L..31H} {939, L31}

\bibitem[\protect\citeauthoryear{{Haslbauer}, {Kroupa}  \& {Jerabkova}}{{Haslbauer} et~al.}{2023}]{2023MNRAS.524.3252H}
{Haslbauer} M.,  {Kroupa} P.,   {Jerabkova} T.,  2023, \mn@doi [\mnras] {10.1093/mnras/stad1986}, \href {https://ui.adsabs.harvard.edu/abs/2023MNRAS.524.3252H} {524, 3252}

\bibitem[\protect\citeauthoryear{{Hopkins}, {Hernquist}, {Cox}, {Di Matteo}, {Robertson}  \& {Springel}}{{Hopkins} et~al.}{2006}]{2006ApJS..163....1H}
{Hopkins} P.~F.,  {Hernquist} L.,  {Cox} T.~J.,  {Di Matteo} T.,  {Robertson} B.,   {Springel} V.,  2006, \mn@doi [\apjs] {10.1086/499298}, \href {https://ui.adsabs.harvard.edu/abs/2006ApJS..163....1H} {163, 1}

\bibitem[\protect\citeauthoryear{{Je{\v{r}}{\'a}bkov{\'a}}, {Kroupa}, {Dabringhausen}, {Hilker}  \& {Bekki}}{{Je{\v{r}}{\'a}bkov{\'a}} et~al.}{2017}]{2017A&A...608A..53J}
{Je{\v{r}}{\'a}bkov{\'a}} T.,  {Kroupa} P.,  {Dabringhausen} J.,  {Hilker} M.,   {Bekki} K.,  2017, \mn@doi [\aap] {10.1051/0004-6361/201731240}, \href {https://ui.adsabs.harvard.edu/abs/2017A&A...608A..53J} {608, A53}

\bibitem[\protect\citeauthoryear{{Je{\v{r}}{\'a}bkov{\'a}}, {Hasani Zonoozi}, {Kroupa}, {Beccari}, {Yan}, {Vazdekis}  \& {Zhang}}{{Je{\v{r}}{\'a}bkov{\'a}} et~al.}{2018}]{2018A&A...620A..39J}
{Je{\v{r}}{\'a}bkov{\'a}} T.,  {Hasani Zonoozi} A.,  {Kroupa} P.,  {Beccari} G.,  {Yan} Z.,  {Vazdekis} A.,   {Zhang} Z.~Y.,  2018, \mn@doi [\aap] {10.1051/0004-6361/201833055}, \href {https://ui.adsabs.harvard.edu/abs/2018A&A...620A..39J} {620, A39}

\bibitem[\protect\citeauthoryear{{Kang} \& {Lee}}{{Kang} \& {Lee}}{2021}]{2021ApJ...914...20K}
{Kang} J.,  {Lee} M.~G.,  2021, \mn@doi [\apj] {10.3847/1538-4357/abf433}, \href {https://ui.adsabs.harvard.edu/abs/2021ApJ...914...20K} {914, 20}

\bibitem[\protect\citeauthoryear{{Katz}, {McGaugh}, {Teuben}  \& {Angus}}{{Katz} et~al.}{2013}]{2013ApJ...772...10K}
{Katz} H.,  {McGaugh} S.,  {Teuben} P.,   {Angus} G.~W.,  2013, \mn@doi [\apj] {10.1088/0004-637X/772/1/10}, \href {https://ui.adsabs.harvard.edu/abs/2013ApJ...772...10K} {772, 10}

\bibitem[\protect\citeauthoryear{{Kroupa}}{{Kroupa}}{2001}]{2001MNRAS.322..231K}
{Kroupa} P.,  2001, \mn@doi [\mnras] {10.1046/j.1365-8711.2001.04022.x}, \href {https://ui.adsabs.harvard.edu/abs/2001MNRAS.322..231K} {322, 231}

\bibitem[\protect\citeauthoryear{{Kroupa} \& {Weidner}}{{Kroupa} \& {Weidner}}{2003}]{2003ApJ...598.1076K}
{Kroupa} P.,  {Weidner} C.,  2003, \mn@doi [\apj] {10.1086/379105}, \href {https://ui.adsabs.harvard.edu/abs/2003ApJ...598.1076K} {598, 1076}

\bibitem[\protect\citeauthoryear{{Kroupa}, {Subr}, {Jerabkova}  \& {Wang}}{{Kroupa} et~al.}{2020}]{2020MNRAS.498.5652K}
{Kroupa} P.,  {Subr} L.,  {Jerabkova} T.,   {Wang} L.,  2020, \mn@doi [\mnras] {10.1093/mnras/staa2276}, \href {https://ui.adsabs.harvard.edu/abs/2020MNRAS.498.5652K} {498, 5652}

\bibitem[\protect\citeauthoryear{{Kroupa} et~al.,}{{Kroupa} et~al.}{2023}]{2023arXiv230911552K}
{Kroupa} P.,  et~al., 2023, \mn@doi [arXiv e-prints] {10.48550/arXiv.2309.11552}, \href {https://ui.adsabs.harvard.edu/abs/2023arXiv230911552K} {p. arXiv:2309.11552}

\bibitem[\protect\citeauthoryear{{Labbe} et~al.,}{{Labbe} et~al.}{2022}]{2022arXiv220712446L}
{Labbe} I.,  et~al., 2022, arXiv e-prints, \href {https://ui.adsabs.harvard.edu/abs/2022arXiv220712446L} {p. arXiv:2207.12446}

\bibitem[\protect\citeauthoryear{{Lelli}, {McGaugh}, {Schombert}  \& {Pawlowski}}{{Lelli} et~al.}{2017}]{2017ApJ...836..152L}
{Lelli} F.,  {McGaugh} S.~S.,  {Schombert} J.~M.,   {Pawlowski} M.~S.,  2017, \mn@doi [\apj] {10.3847/1538-4357/836/2/152}, \href {https://ui.adsabs.harvard.edu/abs/2017ApJ...836..152L} {836, 152}

\bibitem[\protect\citeauthoryear{{L{\"u}ghausen}, {Famaey}  \& {Kroupa}}{{L{\"u}ghausen} et~al.}{2015}]{2015CaJPh..93..232L}
{L{\"u}ghausen} F.,  {Famaey} B.,   {Kroupa} P.,  2015, \mn@doi [Canadian Journal of Physics] {10.1139/cjp-2014-0168}, \href {https://ui.adsabs.harvard.edu/abs/2015CaJPh..93..232L} {93, 232}

\bibitem[\protect\citeauthoryear{{Malekjani}, {Rahvar}  \& {Haghi}}{{Malekjani} et~al.}{2009}]{2009ApJ...694.1220M}
{Malekjani} M.,  {Rahvar} S.,   {Haghi} H.,  2009, \mn@doi [\apj] {10.1088/0004-637X/694/2/1220}, \href {https://ui.adsabs.harvard.edu/abs/2009ApJ...694.1220M} {694, 1220}

\bibitem[\protect\citeauthoryear{{Mart{\'\i}n-Navarro}, {Vazdekis}, {Falc{\'o}n-Barroso}, {La Barbera}, {Y{\i}ld{\i}r{\i}m}  \& {van de Ven}}{{Mart{\'\i}n-Navarro} et~al.}{2018}]{2018MNRAS.475.3700M}
{Mart{\'\i}n-Navarro} I.,  {Vazdekis} A.,  {Falc{\'o}n-Barroso} J.,  {La Barbera} F.,  {Y{\i}ld{\i}r{\i}m} A.,   {van de Ven} G.,  2018, \mn@doi [\mnras] {10.1093/mnras/stx3346}, \href {https://ui.adsabs.harvard.edu/abs/2018MNRAS.475.3700M} {475, 3700}

\bibitem[\protect\citeauthoryear{{Mart{\'\i}n-Navarro}, {van de Ven}  \& {Y{\i}ld{\i}r{\i}m}}{{Mart{\'\i}n-Navarro} et~al.}{2019}]{2019MNRAS.487.4939M}
{Mart{\'\i}n-Navarro} I.,  {van de Ven} G.,   {Y{\i}ld{\i}r{\i}m} A.,  2019, \mn@doi [\mnras] {10.1093/mnras/stz1544}, \href {https://ui.adsabs.harvard.edu/abs/2019MNRAS.487.4939M} {487, 4939}

\bibitem[\protect\citeauthoryear{{McDermid} et~al.,}{{McDermid} et~al.}{2015}]{2015MNRAS.448.3484M}
{McDermid} R.~M.,  et~al., 2015, \mn@doi [\mnras] {10.1093/mnras/stv105}, \href {https://ui.adsabs.harvard.edu/abs/2015MNRAS.448.3484M} {448, 3484}

\bibitem[\protect\citeauthoryear{{McGaugh}}{{McGaugh}}{2004}]{2004ApJ...609..652M}
{McGaugh} S.~S.,  2004, \mn@doi [\apj] {10.1086/421338}, \href {https://ui.adsabs.harvard.edu/abs/2004ApJ...609..652M} {609, 652}

\bibitem[\protect\citeauthoryear{{McGaugh}}{{McGaugh}}{2005}]{2005ApJ...632..859M}
{McGaugh} S.~S.,  2005, \mn@doi [\apj] {10.1086/432968}, \href {https://ui.adsabs.harvard.edu/abs/2005ApJ...632..859M} {632, 859}

\bibitem[\protect\citeauthoryear{{McGaugh}}{{McGaugh}}{2012}]{2012AJ....143...40M}
{McGaugh} S.~S.,  2012, \mn@doi [\aj] {10.1088/0004-6256/143/2/40}, \href {https://ui.adsabs.harvard.edu/abs/2012AJ....143...40M} {143, 40}

\bibitem[\protect\citeauthoryear{{McGaugh}, {Schombert}, {Bothun}  \& {de Blok}}{{McGaugh} et~al.}{2000}]{2000ApJ...533L..99M}
{McGaugh} S.~S.,  {Schombert} J.~M.,  {Bothun} G.~D.,   {de Blok} W.~J.~G.,  2000, \mn@doi [\apjl] {10.1086/312628}, \href {https://ui.adsabs.harvard.edu/abs/2000ApJ...533L..99M} {533, L99}

\bibitem[\protect\citeauthoryear{{Milgrom}}{{Milgrom}}{1983}]{1983ApJ...270..365M}
{Milgrom} M.,  1983, \mn@doi [\apj] {10.1086/161130}, \href {https://ui.adsabs.harvard.edu/abs/1983ApJ...270..365M} {270, 365}

\bibitem[\protect\citeauthoryear{{Milgrom}}{{Milgrom}}{2008}]{2008arXiv0801.3133M}
{Milgrom} M.,  2008, arXiv e-prints, \href {https://ui.adsabs.harvard.edu/abs/2008arXiv0801.3133M} {p. arXiv:0801.3133}

\bibitem[\protect\citeauthoryear{{Milgrom}}{{Milgrom}}{2009}]{2009ApJ...698.1630M}
{Milgrom} M.,  2009, \mn@doi [\apj] {10.1088/0004-637X/698/2/1630}, \href {https://ui.adsabs.harvard.edu/abs/2009ApJ...698.1630M} {698, 1630}

\bibitem[\protect\citeauthoryear{{Milgrom}}{{Milgrom}}{2010}]{2010MNRAS.403..886M}
{Milgrom} M.,  2010, \mn@doi [\mnras] {10.1111/j.1365-2966.2009.16184.x}, \href {https://ui.adsabs.harvard.edu/abs/2010MNRAS.403..886M} {403, 886}

\bibitem[\protect\citeauthoryear{{Milgrom}}{{Milgrom}}{2014}]{2014SchpJ...931410M}
{Milgrom} M.,  2014, \mn@doi [Scholarpedia] {10.4249/scholarpedia.31410}, \href {https://ui.adsabs.harvard.edu/abs/2014SchpJ...931410M} {9, 31410}

\bibitem[\protect\citeauthoryear{{Mo}, {van den Bosch}  \& {White}}{{Mo} et~al.}{2010}]{2010gfe..book.....M}
{Mo} H.,  {van den Bosch} F.~C.,   {White} S.,  2010, {Galaxy Formation and Evolution}

\bibitem[\protect\citeauthoryear{{Mukhanov}}{{Mukhanov}}{2005}]{2005pfc..book.....M}
{Mukhanov} V.,  2005, {Physical Foundations of Cosmology}, \mn@doi{10.2277/0521563984.
}

\bibitem[\protect\citeauthoryear{{Naab}, {Johansson}  \& {Ostriker}}{{Naab} et~al.}{2009}]{2009ApJ...699L.178N}
{Naab} T.,  {Johansson} P.~H.,   {Ostriker} J.~P.,  2009, \mn@doi [\apjl] {10.1088/0004-637X/699/2/L178}, \href {https://ui.adsabs.harvard.edu/abs/2009ApJ...699L.178N} {699, L178}

\bibitem[\protect\citeauthoryear{{Nagesh}, {Banik}, {Thies}, {Kroupa}, {Famaey}, {Wittenburg}, {Parziale}  \& {Haslbauer}}{{Nagesh} et~al.}{2021}]{2021arXiv210111011N}
{Nagesh} S.~T.,  {Banik} I.,  {Thies} I.,  {Kroupa} P.,  {Famaey} B.,  {Wittenburg} N.,  {Parziale} R.,   {Haslbauer} M.,  2021, Canadian Journal of Physics, \href {https://ui.adsabs.harvard.edu/abs/2021arXiv210111011N} {pp 99, 607}

\bibitem[\protect\citeauthoryear{{Nagesh} et~al.,}{{Nagesh} et~al.}{2022}]{2022MNRAS.tmp.3410N}
{Nagesh} S.~T.,  et~al., 2022, \mn@doi [\mnras] {10.1093/mnras/stac3645}, \href {https://ui.adsabs.harvard.edu/abs/2022MNRAS.tmp.3410N} {}

\bibitem[\protect\citeauthoryear{{Naiman} et~al.,}{{Naiman} et~al.}{2018}]{2018MNRAS.477.1206N}
{Naiman} J.~P.,  et~al., 2018, \mn@doi [\mnras] {10.1093/mnras/sty618}, \href {https://ui.adsabs.harvard.edu/abs/2018MNRAS.477.1206N} {477, 1206}

\bibitem[\protect\citeauthoryear{{Nelan}, {Smith}, {Hudson}, {Wegner}, {Lucey}, {Moore}, {Quinney}  \& {Suntzeff}}{{Nelan} et~al.}{2005}]{2005ApJ...632..137N}
{Nelan} J.~E.,  {Smith} R.~J.,  {Hudson} M.~J.,  {Wegner} G.~A.,  {Lucey} J.~R.,  {Moore} S. A.~W.,  {Quinney} S.~J.,   {Suntzeff} N.~B.,  2005, \mn@doi [\apj] {10.1086/431962}, \href {https://ui.adsabs.harvard.edu/abs/2005ApJ...632..137N} {632, 137}

\bibitem[\protect\citeauthoryear{{Nipoti}, {Londrillo}  \& {Ciotti}}{{Nipoti} et~al.}{2007}]{2007ApJ...660..256N}
{Nipoti} C.,  {Londrillo} P.,   {Ciotti} L.,  2007, \mn@doi [\apj] {10.1086/512771}, \href {https://ui.adsabs.harvard.edu/abs/2007ApJ...660..256N} {660, 256}

\bibitem[\protect\citeauthoryear{{Nusser}}{{Nusser}}{2002}]{2002MNRAS.331..909N}
{Nusser} A.,  2002, \mn@doi [\mnras] {10.1046/j.1365-8711.2002.05235.x}, \href {https://ui.adsabs.harvard.edu/abs/2002MNRAS.331..909N} {331, 909}

\bibitem[\protect\citeauthoryear{{Oser}, {Ostriker}, {Naab}, {Johansson}  \& {Burkert}}{{Oser} et~al.}{2010}]{2010ApJ...725.2312O}
{Oser} L.,  {Ostriker} J.~P.,  {Naab} T.,  {Johansson} P.~H.,   {Burkert} A.,  2010, \mn@doi [\apj] {10.1088/0004-637X/725/2/2312}, \href {https://ui.adsabs.harvard.edu/abs/2010ApJ...725.2312O} {725, 2312}

\bibitem[\protect\citeauthoryear{{Peebles}}{{Peebles}}{1980}]{1980lssu.book.....P}
{Peebles} P.~J.~E.,  1980, {The large-scale structure of the universe}

\bibitem[\protect\citeauthoryear{{Pillepich} et~al.,}{{Pillepich} et~al.}{2018}]{Pillepich_2018}
{Pillepich} A.,  et~al., 2018, \mn@doi [\mnras] {10.1093/mnras/stx2656}, \href {https://ui.adsabs.harvard.edu/abs/2018MNRAS.473.4077P} {473, 4077}

\bibitem[\protect\citeauthoryear{{Planck Collaboration} et~al.,}{{Planck Collaboration} et~al.}{2020}]{2020A&A...641A...6P}
{Planck Collaboration} et~al., 2020, \mn@doi [\aap] {10.1051/0004-6361/201833910}, \href {https://ui.adsabs.harvard.edu/abs/2020A&A...641A...6P} {641, A6}

\bibitem[\protect\citeauthoryear{{Ploeckinger}, {Schaye}, {Hacar}, {Maseda}, {Hodge}  \& {Bouwens}}{{Ploeckinger} et~al.}{2019}]{2019MNRAS.484.4379P}
{Ploeckinger} S.,  {Schaye} J.,  {Hacar} A.,  {Maseda} M.~V.,  {Hodge} J.~A.,   {Bouwens} R.~J.,  2019, \mn@doi [\mnras] {10.1093/mnras/stz173}, \href {https://ui.adsabs.harvard.edu/abs/2019MNRAS.484.4379P} {484, 4379}

\bibitem[\protect\citeauthoryear{{Rodriguez-Gomez} et~al.,}{{Rodriguez-Gomez} et~al.}{2016}]{2016MNRAS.458.2371R}
{Rodriguez-Gomez} V.,  et~al., 2016, \mn@doi [\mnras] {10.1093/mnras/stw456}, \href {https://ui.adsabs.harvard.edu/abs/2016MNRAS.458.2371R} {458, 2371}

\bibitem[\protect\citeauthoryear{{Romanowsky} \& {Fall}}{{Romanowsky} \& {Fall}}{2012}]{2012ApJS..203...17R}
{Romanowsky} A.~J.,  {Fall} S.~M.,  2012, \mn@doi [\apjs] {10.1088/0067-0049/203/2/17}, \href {https://ui.adsabs.harvard.edu/abs/2012ApJS..203...17R} {203, 17}

\bibitem[\protect\citeauthoryear{{Salvador-Rusi{\~n}ol}, {Vazdekis}, {La Barbera}, {Beasley}, {Ferreras}, {Negri}  \& {Dalla Vecchia}}{{Salvador-Rusi{\~n}ol} et~al.}{2020}]{2020NatAs...4..252S}
{Salvador-Rusi{\~n}ol} N.,  {Vazdekis} A.,  {La Barbera} F.,  {Beasley} M.~A.,  {Ferreras} I.,  {Negri} A.,   {Dalla Vecchia} C.,  2020, \mn@doi [Nature Astronomy] {10.1038/s41550-019-0955-0}, \href {https://ui.adsabs.harvard.edu/abs/2020NatAs...4..252S} {4, 252}

\bibitem[\protect\citeauthoryear{{Salvador-Rusi{\~n}ol}, {Beasley}, {Vazdekis}  \& {Barbera}}{{Salvador-Rusi{\~n}ol} et~al.}{2021}]{2021MNRAS.500.3368S}
{Salvador-Rusi{\~n}ol} N.,  {Beasley} M.~A.,  {Vazdekis} A.,   {Barbera} F.~L.,  2021, \mn@doi [\mnras] {10.1093/mnras/staa3419}, \href {https://ui.adsabs.harvard.edu/abs/2021MNRAS.500.3368S} {500, 3368}

\bibitem[\protect\citeauthoryear{{Salvador-Rusi{\~n}ol}, {Ferr{\'e}-Mateu}, {Vazdekis}  \& {Beasley}}{{Salvador-Rusi{\~n}ol} et~al.}{2022}]{2022MNRAS.515.4514S}
{Salvador-Rusi{\~n}ol} N.,  {Ferr{\'e}-Mateu} A.,  {Vazdekis} A.,   {Beasley} M.~A.,  2022, \mn@doi [\mnras] {10.1093/mnras/stac2070}, \href {https://ui.adsabs.harvard.edu/abs/2022MNRAS.515.4514S} {515, 4514}

\bibitem[\protect\citeauthoryear{{Sanders}}{{Sanders}}{1990}]{1990A&ARv...2....1S}
{Sanders} R.~H.,  1990, \mn@doi [\aapr] {10.1007/BF00873540}, \href {https://ui.adsabs.harvard.edu/abs/1990A&ARv...2....1S} {2, 1}

\bibitem[\protect\citeauthoryear{{Sanders}}{{Sanders}}{1998}]{1998MNRAS.296.1009S}
{Sanders} R.~H.,  1998, \mn@doi [\mnras] {10.1046/j.1365-8711.1998.01459.x}, \href {https://ui.adsabs.harvard.edu/abs/1998MNRAS.296.1009S} {296, 1009}

\bibitem[\protect\citeauthoryear{{Sanders}}{{Sanders}}{2008}]{2008MNRAS.386.1588S}
{Sanders} R.~H.,  2008, \mn@doi [\mnras] {10.1111/j.1365-2966.2008.13140.x}, \href {https://ui.adsabs.harvard.edu/abs/2008MNRAS.386.1588S} {386, 1588}

\bibitem[\protect\citeauthoryear{{Schaye} et~al.,}{{Schaye} et~al.}{2015}]{2015MNRAS.446..521S}
{Schaye} J.,  et~al., 2015, \mn@doi [\mnras] {10.1093/mnras/stu2058}, \href {https://ui.adsabs.harvard.edu/abs/2015MNRAS.446..521S} {446, 521}

\bibitem[\protect\citeauthoryear{{Skordis} \& {Z{\l}o{\'s}nik}}{{Skordis} \& {Z{\l}o{\'s}nik}}{2021}]{2021PhRvL.127p1302S}
{Skordis} C.,  {Z{\l}o{\'s}nik} T.,  2021, \mn@doi [\prl] {10.1103/PhysRevLett.127.161302}, \href {https://ui.adsabs.harvard.edu/abs/2021PhRvL.127p1302S} {127, 161302}

\bibitem[\protect\citeauthoryear{{Somerville} \& {Dav{\'e}}}{{Somerville} \& {Dav{\'e}}}{2015}]{2015ARA&A..53...51S}
{Somerville} R.~S.,  {Dav{\'e}} R.,  2015, \mn@doi [\araa] {10.1146/annurev-astro-082812-140951}, \href {https://ui.adsabs.harvard.edu/abs/2015ARA&A..53...51S} {53, 51}

\bibitem[\protect\citeauthoryear{{Spiniello} et~al.,}{{Spiniello} et~al.}{2021}]{2021A&A...654A.136S}
{Spiniello} C.,  et~al., 2021, \mn@doi [\aap] {10.1051/0004-6361/202140856}, \href {https://ui.adsabs.harvard.edu/abs/2021A&A...654A.136S} {654, A136}

\bibitem[\protect\citeauthoryear{{Spiniello} et~al.,}{{Spiniello} et~al.}{2024}]{2024MNRAS.527.8793S}
{Spiniello} C.,  et~al., 2024, \mn@doi [\mnras] {10.1093/mnras/stad3703}, \href {https://ui.adsabs.harvard.edu/abs/2024MNRAS.527.8793S} {527, 8793}

\bibitem[\protect\citeauthoryear{{Stringer}, {Trujillo}, {Dalla Vecchia}  \& {Martinez-Valpuesta}}{{Stringer} et~al.}{2015}]{2015MNRAS.449.2396S}
{Stringer} M.,  {Trujillo} I.,  {Dalla Vecchia} C.,   {Martinez-Valpuesta} I.,  2015, \mn@doi [\mnras] {10.1093/mnras/stv455}, \href {https://ui.adsabs.harvard.edu/abs/2015MNRAS.449.2396S} {449, 2396}

\bibitem[\protect\citeauthoryear{{Tamburri}, {Saracco}, {Longhetti}, {Gargiulo}, {Lonoce}  \& {Ciocca}}{{Tamburri} et~al.}{2014}]{2014A&A...570A.102T}
{Tamburri} S.,  {Saracco} P.,  {Longhetti} M.,  {Gargiulo} A.,  {Lonoce} I.,   {Ciocca} F.,  2014, \mn@doi [\aap] {10.1051/0004-6361/201424040}, \href {https://ui.adsabs.harvard.edu/abs/2014A&A...570A.102T} {570, A102}

\bibitem[\protect\citeauthoryear{{Teyssier}}{{Teyssier}}{2002}]{2002A&A...385..337T}
{Teyssier} R.,  2002, \mn@doi [\aap] {10.1051/0004-6361:20011817}, \href {https://ui.adsabs.harvard.edu/abs/2002A&A...385..337T} {385, 337}

\bibitem[\protect\citeauthoryear{{Thomas}, {Maraston}, {Bender}  \& {Mendes de Oliveira}}{{Thomas} et~al.}{2005}]{2005ApJ...621..673T}
{Thomas} D.,  {Maraston} C.,  {Bender} R.,   {Mendes de Oliveira} C.,  2005, \mn@doi [\apj] {10.1086/426932}, \href {https://ui.adsabs.harvard.edu/abs/2005ApJ...621..673T} {621, 673}

\bibitem[\protect\citeauthoryear{{Thomas}, {Maraston}, {Schawinski}, {Sarzi}  \& {Silk}}{{Thomas} et~al.}{2010}]{2010MNRAS.404.1775T}
{Thomas} D.,  {Maraston} C.,  {Schawinski} K.,  {Sarzi} M.,   {Silk} J.,  2010, \mn@doi [\mnras] {10.1111/j.1365-2966.2010.16427.x}, \href {https://ui.adsabs.harvard.edu/abs/2010MNRAS.404.1775T} {404, 1775}

\bibitem[\protect\citeauthoryear{{Trujillo}, {Cenarro}, {de Lorenzo-C{\'a}ceres}, {Vazdekis}, {de la Rosa}  \& {Cava}}{{Trujillo} et~al.}{2009}]{2009ApJ...692L.118T}
{Trujillo} I.,  {Cenarro} A.~J.,  {de Lorenzo-C{\'a}ceres} A.,  {Vazdekis} A.,  {de la Rosa} I.~G.,   {Cava} A.,  2009, \mn@doi [\apjl] {10.1088/0004-637X/692/2/L118}, \href {https://ui.adsabs.harvard.edu/abs/2009ApJ...692L.118T} {692, L118}

\bibitem[\protect\citeauthoryear{{Trujillo}, {Carrasco}  \& {Ferr{\'e}-Mateu}}{{Trujillo} et~al.}{2012}]{2012ApJ...751...45T}
{Trujillo} I.,  {Carrasco} E.~R.,   {Ferr{\'e}-Mateu} A.,  2012, \mn@doi [\apj] {10.1088/0004-637X/751/1/45}, \href {https://ui.adsabs.harvard.edu/abs/2012ApJ...751...45T} {751, 45}

\bibitem[\protect\citeauthoryear{{Trujillo}, {Ferr{\'e}-Mateu}, {Balcells}, {Vazdekis}  \& {S{\'a}nchez-Bl{\'a}zquez}}{{Trujillo} et~al.}{2014}]{2014ApJ...780L..20T}
{Trujillo} I.,  {Ferr{\'e}-Mateu} A.,  {Balcells} M.,  {Vazdekis} A.,   {S{\'a}nchez-Bl{\'a}zquez} P.,  2014, \mn@doi [\apjl] {10.1088/2041-8205/780/2/L20}, \href {https://ui.adsabs.harvard.edu/abs/2014ApJ...780L..20T} {780, L20}

\bibitem[\protect\citeauthoryear{{Vazdekis}, {Koleva}, {Ricciardelli}, {R{\"o}ck}  \& {Falc{\'o}n-Barroso}}{{Vazdekis} et~al.}{2016}]{2016MNRAS.463.3409V}
{Vazdekis} A.,  {Koleva} M.,  {Ricciardelli} E.,  {R{\"o}ck} B.,   {Falc{\'o}n-Barroso} J.,  2016, \mn@doi [\mnras] {10.1093/mnras/stw2231}, \href {https://ui.adsabs.harvard.edu/abs/2016MNRAS.463.3409V} {463, 3409}

\bibitem[\protect\citeauthoryear{{Vogelsberger}, {Marinacci}, {Torrey}  \& {Puchwein}}{{Vogelsberger} et~al.}{2020}]{2020NatRP...2...42V}
{Vogelsberger} M.,  {Marinacci} F.,  {Torrey} P.,   {Puchwein} E.,  2020, \mn@doi [Nature Reviews Physics] {10.1038/s42254-019-0127-2}, \href {https://ui.adsabs.harvard.edu/abs/2020NatRP...2...42V} {2, 42}

\bibitem[\protect\citeauthoryear{{Walsh}, {van den Bosch}, {Gebhardt}, {Y{\i}ld{\i}r{\i}m}, {Richstone}, {G{\"u}ltekin}  \& {Husemann}}{{Walsh} et~al.}{2016}]{2016ApJ...817....2W}
{Walsh} J.~L.,  {van den Bosch} R. C.~E.,  {Gebhardt} K.,  {Y{\i}ld{\i}r{\i}m} A.,  {Richstone} D.~O.,  {G{\"u}ltekin} K.,   {Husemann} B.,  2016, \mn@doi [\apj] {10.3847/0004-637X/817/1/2}, \href {https://ui.adsabs.harvard.edu/abs/2016ApJ...817....2W} {817, 2}

\bibitem[\protect\citeauthoryear{{Weidner} \& {Kroupa}}{{Weidner} \& {Kroupa}}{2006}]{2006MNRAS.365.1333W}
{Weidner} C.,  {Kroupa} P.,  2006, \mn@doi [\mnras] {10.1111/j.1365-2966.2005.09824.x}, \href {https://ui.adsabs.harvard.edu/abs/2006MNRAS.365.1333W} {365, 1333}

\bibitem[\protect\citeauthoryear{{Wellons} et~al.,}{{Wellons} et~al.}{2015}]{2015MNRAS.449..361W}
{Wellons} S.,  et~al., 2015, \mn@doi [\mnras] {10.1093/mnras/stv303}, \href {https://ui.adsabs.harvard.edu/abs/2015MNRAS.449..361W} {449, 361}

\bibitem[\protect\citeauthoryear{{Wellons} et~al.,}{{Wellons} et~al.}{2016}]{2016MNRAS.456.1030W}
{Wellons} S.,  et~al., 2016, \mn@doi [\mnras] {10.1093/mnras/stv2738}, \href {https://ui.adsabs.harvard.edu/abs/2016MNRAS.456.1030W} {456, 1030}

\bibitem[\protect\citeauthoryear{{Wittenburg}, {Kroupa}  \& {Famaey}}{{Wittenburg} et~al.}{2020}]{2020ApJ...890..173W}
{Wittenburg} N.,  {Kroupa} P.,   {Famaey} B.,  2020, \mn@doi [\apj] {10.3847/1538-4357/ab6d73}, \href {https://ui.adsabs.harvard.edu/abs/2020ApJ...890..173W} {890, 173}

\bibitem[\protect\citeauthoryear{{Wittenburg}, {Kroupa}, {Banik}, {Candlish}  \& {Samaras}}{{Wittenburg} et~al.}{2023}]{2023MNRAS.523..453W}
{Wittenburg} N.,  {Kroupa} P.,  {Banik} I.,  {Candlish} G.,   {Samaras} N.,  2023, \mn@doi [\mnras] {10.1093/mnras/stad1371}, \href {https://ui.adsabs.harvard.edu/abs/2023MNRAS.523..453W} {523, 453}

\bibitem[\protect\citeauthoryear{{Yan}, {Je{\v{r}}{\'a}bkov{\'a}}  \& {Kroupa}}{{Yan} et~al.}{2021}]{2021A&A...655A..19Y}
{Yan} Z.,  {Je{\v{r}}{\'a}bkov{\'a}} T.,   {Kroupa} P.,  2021, \mn@doi [\aap] {10.1051/0004-6361/202140683}, \href {https://ui.adsabs.harvard.edu/abs/2021A&A...655A..19Y} {655, A19}

\bibitem[\protect\citeauthoryear{{Y{\i}ld{\i}r{\i}m}, {van den Bosch}, {van de Ven}, {Husemann}, {Lyubenova}, {Walsh}, {Gebhardt}  \& {G{\"u}ltekin}}{{Y{\i}ld{\i}r{\i}m} et~al.}{2015}]{2015MNRAS.452.1792Y}
{Y{\i}ld{\i}r{\i}m} A.,  {van den Bosch} R. C.~E.,  {van de Ven} G.,  {Husemann} B.,  {Lyubenova} M.,  {Walsh} J.~L.,  {Gebhardt} K.,   {G{\"u}ltekin} K.,  2015, \mn@doi [\mnras] {10.1093/mnras/stv1381}, \href {https://ui.adsabs.harvard.edu/abs/2015MNRAS.452.1792Y} {452, 1792}

\bibitem[\protect\citeauthoryear{{Y{\i}ld{\i}r{\i}m}, {van den Bosch}, {van de Ven}, {Mart{\'\i}n-Navarro}, {Walsh}, {Husemann}, {G{\"u}ltekin}  \& {Gebhardt}}{{Y{\i}ld{\i}r{\i}m} et~al.}{2017}]{2017MNRAS.468.4216Y}
{Y{\i}ld{\i}r{\i}m} A.,  {van den Bosch} R. C.~E.,  {van de Ven} G.,  {Mart{\'\i}n-Navarro} I.,  {Walsh} J.~L.,  {Husemann} B.,  {G{\"u}ltekin} K.,   {Gebhardt} K.,  2017, \mn@doi [\mnras] {10.1093/mnras/stx732}, \href {https://ui.adsabs.harvard.edu/abs/2017MNRAS.468.4216Y} {468, 4216}

\bibitem[\protect\citeauthoryear{{de La Rosa}, {La Barbera}, {Ferreras}  \& {de Carvalho}}{{de La Rosa} et~al.}{2011}]{2011MNRAS.418L..74D}
{de La Rosa} I.~G.,  {La Barbera} F.,  {Ferreras} I.,   {de Carvalho} R.~R.,  2011, \mn@doi [\mnras] {10.1111/j.1745-3933.2011.01146.x}, \href {https://ui.adsabs.harvard.edu/abs/2011MNRAS.418L..74D} {418, L74}

\bibitem[\protect\citeauthoryear{{van Dokkum} et~al.,}{{van Dokkum} et~al.}{2010}]{2010ApJ...709.1018V}
{van Dokkum} P.~G.,  et~al., 2010, \mn@doi [\apj] {10.1088/0004-637X/709/2/1018}, \href {https://ui.adsabs.harvard.edu/abs/2010ApJ...709.1018V} {709, 1018}

\bibitem[\protect\citeauthoryear{{van Son} et~al.,}{{van Son} et~al.}{2019}]{2019MNRAS.485..396V}
{van Son} L. A.~C.,  et~al., 2019, \mn@doi [\mnras] {10.1093/mnras/stz399}, \href {https://ui.adsabs.harvard.edu/abs/2019MNRAS.485..396V} {485, 396}

\bibitem[\protect\citeauthoryear{{van den Bosch}, {Gebhardt}, {G{\"u}ltekin}, {van de Ven}, {van der Wel}  \& {Walsh}}{{van den Bosch} et~al.}{2012}]{2012Natur.491..729V}
{van den Bosch} R. C.~E.,  {Gebhardt} K.,  {G{\"u}ltekin} K.,  {van de Ven} G.,  {van der Wel} A.,   {Walsh} J.~L.,  2012, \mn@doi [\nat] {10.1038/nature11592}, \href {https://ui.adsabs.harvard.edu/abs/2012Natur.491..729V} {491, 729}

\bibitem[\protect\citeauthoryear{{van der Wel} et~al.,}{{van der Wel} et~al.}{2014}]{2014ApJ...788...28V}
{van der Wel} A.,  et~al., 2014, \mn@doi [\apj] {10.1088/0004-637X/788/1/28}, \href {https://ui.adsabs.harvard.edu/abs/2014ApJ...788...28V} {788, 28}

\makeatother
\end{thebibliography}



\appendix

\section{Specific angular momentum}
\label{sec:appendix2}
The specific angular momentum vector is obtained from,
\begin{equation}
\mathbf{j} = \frac{\int_{r} \mathbf{{r_{i}}} \times \mathbf{{v_{i}}} \ m_{i} \ d^{3}\mathbf{{r}}}{\int_{r} m_{i} \ d^{3}\mathbf{{r}}},
	\label{eq:angmomvec}
\end{equation}
where $\mathbf{{r_{i}}}$, $\mathbf{{v_{i}}}$ and $m_{i}$ are respectively the position- and velocity-vectors and mass of the stellar particle $i$. The absolute value of the specific angular momentum is,
\begin{equation}
\begin{split}
j_{\star} = |\mathbf{j}| = (j_{\rm x}^{2}+j_{\rm y}^{2}+j_{\rm z}^{2})^{0.5},
\end{split}
	\label{eq:spangmom}
\end{equation}
where $j_{\rm x}$, $j_{\rm y}$ and $j_{\rm z}$ are the x, y and z components of $\mathbf{j}$.


\section{Calculation of velocity dispersion}
\label{sec:appendix1}
The mass-weighted square of the line-of-sight velocity dispersion for the model galaxy, $\sigma^{2}$, is the statistical dispersion of speeds about the mean speed $\overline{v}$ at each radius,
\begin{equation}
\mathbf{\sigma}^{2} = \left(\frac{\Sigma_{i}^{n} (m_{i} \cdot v_{i} - m_{\rm total} \cdot \overline{v})^{2}}{m_{\rm total}}\right)^{0.5}
	\label{eq:vdisp}
\end{equation}
where $m_{\rm i}$ is the mass of the stellar particle $i$, $v_{\rm i}$ is the line of sight speed of the stellar particle and $m_{\rm total}$ is the total mass of stellar particles within the radius $r$.


\bsp	
\label{lastpage}
\end{document}